\documentclass[]{aa}  

\usepackage{graphicx}
\usepackage{multirow}
\usepackage{txfonts}
\begin{document}

   \title{High turbulence in the IM Lup protoplanetary disk}
    \subtitle{Direct observational constraints from CN and C$_2$H emission}
   \authorrunning{T. Paneque-Carre\~no et al.}
    \author{T. Paneque-Carre\~no \inst{1,2},
            A. F. Izquierdo\inst{1,2},
            R. Teague \inst{3},
            A. Miotello \inst{1},
            E. A. Bergin \inst{4},
            R. Loomis \inst{5},
            E. F. van Dishoeck \inst{2,6}}
          
    \institute{European Southern Observatory, Karl-Schwarzschild-Str 2, 85748 Garching, Germany
    \and Leiden Observatory, Leiden University, P.O. Box 9513, NL-2300 RA Leiden, the Netherlands
    \and Department of Earth, Atmospheric, and Planetary Sciences, Massachusetts Institute of Technology, Cambridge, MA 02139, USA
    \and Department of Astronomy, University of Michigan, 323 West Hall, 1085 S. University Avenue, Ann Arbor, MI 48109, USA
    \and National Radio Astronomy Observatory, Charlottesville, VA 22903, USA
    \and Max-Planck-Institut für extraterrestrische Physik, Gießenbachstr. 1 , 85748 Garching bei München, Germany
    \\
              \email{tpaneque@eso.org}}
   \date{}

% \abstract{}{}{}{}{} 
% 5 {} token are mandatory
 
  \abstract
  % context heading (optional)
  % {} leave it empty if necessary  
   {Constraining turbulence in disks is key to understanding their evolution via the transport of angular momentum. Measurements of high turbulence remain elusive, and methods for estimating turbulence mostly rely on complex radiative transfer models of the data. Using the disk emission from IM Lup, a source proposed to be undergoing magneto-rotational instabilities (MRIs) and to possibly have high turbulence values in the upper disk layers, we present a new way of directly measuring turbulence without the need of radiative transfer or thermochemical models.}
  % aims heading (mandatory)
   {Through the characterization of the CN and C$_2$H emission in IM Lup, we aim to connect the information on the vertical and thermal structure of a particular disk region to derive the turbulence at that location. By using an optically thin tracer, it is possible to directly measure turbulence from the nonthermal broadening of the line.}
  % methods heading (mandatory)
   {The vertical layers of the CN and C$_2$H emission were traced directly from the channel maps using ALFAHOR. By comparing their position to that of optically thick CO observations, we were able to characterize the kinetic temperature of the emitting region. Using a simple parametric model of the line intensity with DISCMINER, we accurately measured the emission linewidth and separated the thermal and nonthermal components. Assuming that the nonthermal component is fully turbulent, we were able to directly estimate the turbulent motions at the studied radial and vertical location of CN emission.}
  % results heading (mandatory)
   {IM Lup shows a high turbulence of Mach 0.4-0.6 at $z/r \sim$ 0.25. Considering previous estimates of low turbulence near the midplane, this may indicate a vertical gradient in the disk turbulence, which is a key prediction in MRI studies. CN and C$_2$H are both emitting from a localized upper disk region at $z/r =$0.2-0.3, in agreement with thermochemical models.}
  % conclusions heading (optional), leave it empty if necessary 
   {}


   \maketitle
%
%-------------------------------------------------------------------

\section{Introduction}

The study of planet formation requires high sensitivity observations that can spatially and spectrally resolve the motions of the material in protoplanetary disks around young stars. In this context, the Atacama Large Millimeter/submillimeter Array (ALMA) is a unique facility, able to characterize these objects and their evolution mechanisms, which depend on the transportation of angular momentum \citep{Lynden_Bell_Pringle_1974}. Turbulence is the mechanism proposed to shape the surface density of disks, by transporting angular momentum through velocity fluctuations and providing an effective viscosity at microscopic scales \citep{Shakura_Sunyaev_1973_visc}. There are many hydrodynamic processes that could be the origin of turbulent motions \citep[see][and references therein]{Lesur_ppvii_2023}; however, direct measurements of turbulence in protoplanetary disks have been hard to obtain, with only a few upper limits \citep[e.g.,][]{Hughes_2009_turb, Flaherty_2015_hd16_weakturb, Teague_2016_turb_TWHya} and one candidate with measured turbulence, DM Tau. In this system, high spatial and spectral resolution ALMA observations of CS allowed \citet{Guilloteau_2012_turb_dmtau} to initially measure turbulence values of 0.3-0.5$c_s$ (see also the initial studies by \citealt{Guilloteau_Dutrey_1998} and \citealt{Dartois_2003}). A later study by \citet{Flaherty_2020_DMTau} obtained lower, but still significant, turbulence values of 0.25-0.33$c_s$ through modeling of CO emission in DM Tau. 

Current methods used to constrain turbulence mostly depend on comparing observations to complex parametric and radiative transfer models, which depend on fundamental physical disk properties such as surface density and temperature structure \citep[e.g.,][]{Hughes_2009_turb, Guilloteau_2012_turb_dmtau, Flaherty_2015_hd16_weakturb, Flaherty_2017_DCOp, Flaherty_2020_DMTau}. While these methods constrain their best-fit parameters with state-of-the-art analysis, there are multiple effects that are hard to account for since precisely resolving the gas surface density profile requires knowledge of the dust grain distribution and excitation conditions of the observed molecule \citep{MAPS_Zhang, Miotello_ppvii_2023}. Furthermore, the presence of substructure in the disk may or may not cause gaps and rings in the surface density, depending on their origin, which is also not trivial to determine \citep{Rosotti_2021_CITau, MAPS_Zhang, Teague_2018_hd16planet}.

Due to their complexity, previous uses of these models to extract disk turbulence have assumed a single turbulence value for the whole protoplanetary disk, neglecting the possibility of a radial or vertical dependence. However, turbulence could have vertical variations, depending on the launching mechanism \citep{Forgan_2012, Simon_2015, shi_chiang_2014}, and while various molecular tracers are used to probe the vertical structure in searches for vertical turbulence variations \citep{Flaherty_2017_DCOp}, locating the exact position of the molecular emission in the disk is not straightforward in the modeling process and in some cases presents strong discrepancies with observational constraints. A successful case is the best-fit model to extract turbulence in DM Tau, where the CO ($J = 2-1$) emission is expected to arise from a layer of $z/r \sim$0.3-0.4 \citep[see Figure 2 in][]{Flaherty_2020_DMTau}, in good agreement with observational constraints from \citet{Law_2023_12CO_surf}. However, analysis of HD\,163296, where no significant turbulence was detected, predicts that CO ($J = 2-1$) emission is coming from $z/r <$0.1 \citep[see Figure 16 in][]{Flaherty_2017_DCOp}, which is many times lower than the measured emission surface of CO ($J = 2-1$), located at $z/r \sim$0.3 \citep{MAPS_Law_Surf, Paneque_2023_vert}. These differences may be due to the limitations of the method, a lack of spatial resolution, or errors in the assumed density and temperature structure.

%-chemistry of CN and C2H, emission constrained to thin layer.
To alleviate these issues, this work focuses on directly tracing the vertical location of optically thin CN and C$_2$H transitions, measuring the radial profile of line broadening due to nonthermal motions, and relating this to vertically and radially resolved turbulence. CN and C$_2$H are expected to originate from chemical processes requiring energetic UV-photon radiation \citep{Visser_2018, Cazzoletti_2018_CN, Bergin_2016_c2h, Teague_2020_CN}, and therefore their emission should arise from a sweet spot located at high elevation from the midplane \citep{Schreyer_2008, Cazzoletti_2018_CN}, in a zone known as the photon-dominated region \citep{Aikawa_2002}. As the emission should come from a vertically thin slab, it is possible to accurately determine the location of the emission surfaces of CN and C$_2$H despite the emission being optically thin \citep[see][]{Paneque-Carreno_2022_Elias_CN}. In parallel, optically thick CO isotopologs can be used as tracers of the disk structure, with the resolved vertical emission location \citep{MAPS_Law_Surf, Paneque-Carreno_2022_Elias_CN} allowing us to provide a global three-dimensional characterization of the temperature structure, molecular emission, and turbulence in the upper disk layers. This is ultimately achieved by measuring the emission linewidth and separating the thermal from the nonthermal component. As we assume optically thin molecular emission, the nonthermal component is assumed to be mainly turbulent \citep{Hacar_2016_optdepth}. A key tool in our analysis is the use of  DISCMINER \citep{izquierdo_2021_discminer1, Izquierdo_2023_discminer2}, a disk analysis package that fits the spectra per pixel of our data to precisely extract and parametrize the radial linewidth profile of the data, accounting for the effects of beam convolution and upper and lower emission surfaces.

%-IM Lup target of opportunity

Our work focuses on the protoplanetary disk around the young K5 star IM Lup, located at a distance of 158\,pc \citep{Gaia_DR3}. This system has been extensively studied using multiple tracers. Infrared scattered light shows a large, vertically extended disk, with radial rings and gaps \citep{Avenhaus_2018}. Millimeter continuum traces tightly wound spiral arms and an outer ring \citep{Cleeves_LupusModel_2016, DSHARP_Huang_radial, DSHARP_Huang_Spirals}, and gas line emission shows a rich molecular reservoir \citep{Pinte_2018_method, MAPS_Oberg}. Recent observational and theoretical works have proposed that IM Lup has an inner turbulent disk \citep{Bosman_2023_imlup} and possible turbulent velocities of $\sim$100\,m\,s$^{-1}$ in the outer disk \citep{Cleeves_LupusModel_2016}. Due to its inclination and vertical extent, it is possible to trace the vertical disk structure and molecular layering of various lines; however, previous studies have been unable to extract CN or C$_2$H vertical profiles due the low signal-to-noise ratio (S/N) of the lower rotational transitions \citep{Paneque_2023_vert}. New, higher-energy ALMA Band 7 transitions have sufficient S/N for an adequate characterization of their radial, azimuthal, and vertical features.

This paper is structured as follows. Section 2 presents the data and explains the calibration procedures. Section 3 studies the emission morphology, characterizing the vertical extent of the disk and the brightness temperature of CN and C$_2$H and comparing them to those of CO isotopologs. Section 4 shows the parametric model we used to extract the linewidth from CN, and Section 5 presents our results on the radial turbulence profile as traced by CN. Finally, Section 6 discusses the findings of this work in the context of previous constrains, and Section 7 summarizes our main conclusions.

\begin{figure*}[h!]
   \centering
   \includegraphics[width=\hsize]{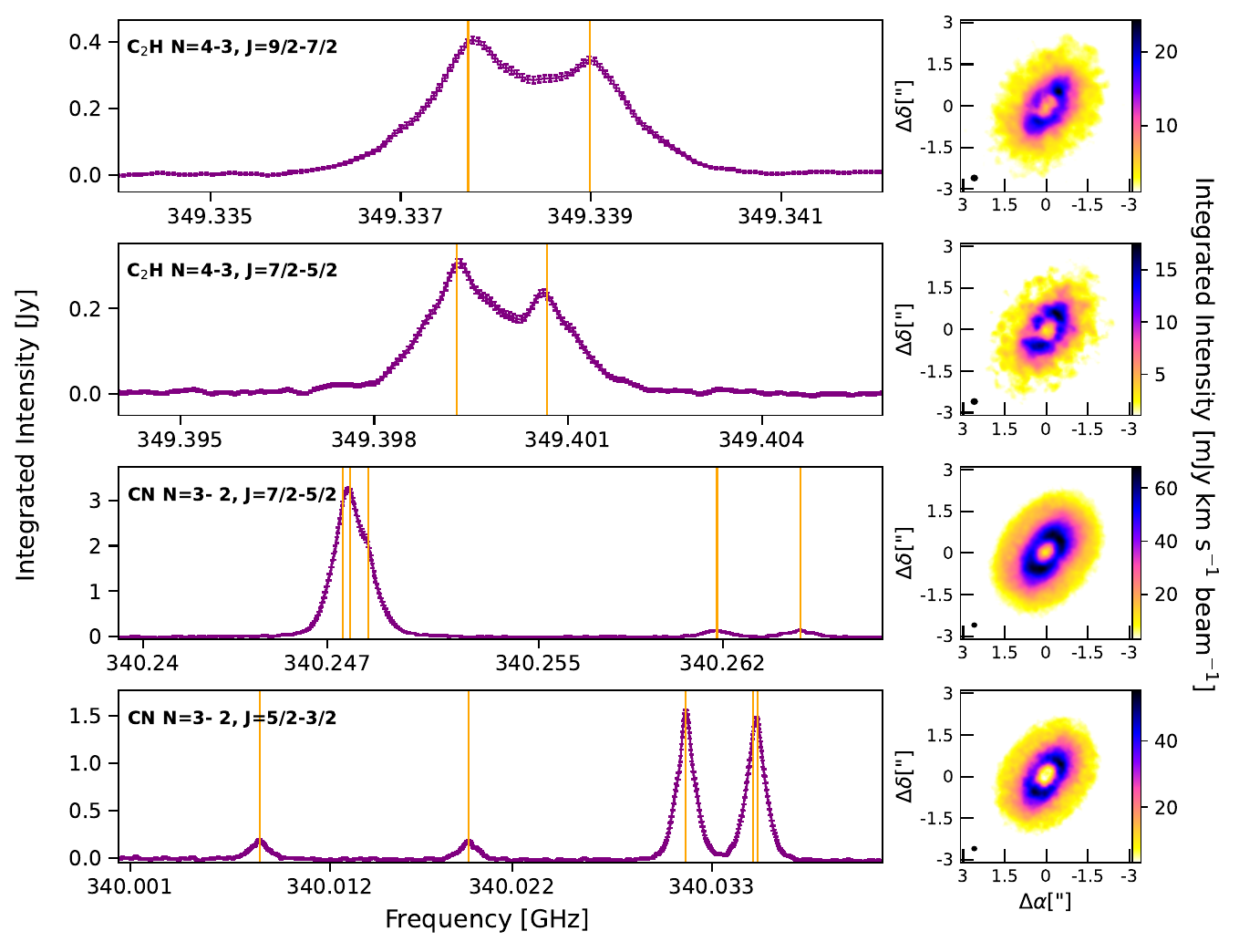}
      \caption{ALMA observations of CN and C$_2$H line emission from IM Lup. Left column: Stacked, integrated spectra extracted from the whole radial extent of the disk, for each transition, obtained using GoFish. The corresponding hyperfine groups are indicated in the top left of each panel. Vertical lines mark the location of the central frequency for each hyperfine component within the spectral range. The channel spacing is 50\,m\,s$^{-1}$, matching the native spectral resolution of the data.  Right column: Integrated intensity maps (moment 0). 
              }
         \label{imlup_emission}
\end{figure*}

\begin{table*}[h]
\def\arraystretch{1.5}
\setlength{\tabcolsep}{7pt}
\caption{Molecular line data from the LAMBDA database \citep{lambda_database_Schoier_2005}.}
\label{table_molec}      
\centering
\begin{tabular}{c| c| c c c c c c c}       
\hline\hline                
Species & $J$ Transition & $F$ Transition& Frequency & $E_u$ &  $A_{ul}$ & $g_u$ & Int. Flux \tablefootmark{*}\\   
 &  & & (GHz)& (K) & (s$^{-1}$ )& & (Jy\,km\,s$^{-1}$)\\   
\hline                       

   \multirow{10}{*}{CN ($N=3-2$)} & \multirow{5}{*}{$J=7/2-5/2$}  &  $F=7/2-5/2$ &340.24759\tablefootmark{a}  & 32.67 &3.797$\times$10$^{-4}$   & 8 & \multirow{3}{*}{4.710 $\pm$ 0.029 }\\ 
    &   & $F=9/2-7/2$ & 340.24786\tablefootmark{a} & 32.67 &4.13$\times$10$^{-4}$  & 10 &\\ 
     &   & $F=5/2-3/2$ & 340.24854 &32.67 &3.67$\times$10$^{-4}$  & 6 & \\ 
\cline{3-8}
    &   & $F=5/2-5/2$ & 340.26177 &32.67 &0.45$\times$10$^{-4}$  & 6 &  0.133 $\pm$ 0.029\\ 
\cline{3-8}
    &   & $F=7/2-7/2$ & 340.26495 &32.67 &0.34$\times$10$^{-4}$  & 8 & 0.137 $\pm$ 0.029\\ 
\cline{2-8}
   & \multirow{5}{*}{$J=5/2-3/2$} & $F=5/2-5/2$ & 340.00813  &  32.63 &0.62$\times$10$^{-4}$ & 6 & 0.140 $\pm$ 0.001\\
\cline{3-8}  
    & & $F=3/2-3/2$ & 340.01963  &  32.63 &0.93$\times$10$^{-4}$ & 4 & 0.091 $\pm$ 0.001\\
\cline{3-8}  
   & & $F=7/2-5/2$ & 340.03155  &  32.63 &3.84$\times$10$^{-4}$ & 8 & 1.594 $\pm$ 0.001\\
\cline{3-8}  
   & & $F=3/2-1/2$&  340.03527\tablefootmark{b} & 32.63 &2.89$\times$10$^{-4}$  & 4 & \multirow{2}{*}{1.610 $\pm$ 0.001}\\
   &  & $F=5/2-3/2$&  340.03551\tablefootmark{b}  &  32.63 &3.23$\times$10$^{-4}$  & 6 & \\
\hline

\multirow{4}{*}{C$_2$H ($N=4-3$)} & \multirow{2}{*}{$J=9/2-7/2$} & $F = 5-4$ & 349.33771 & 41.91 & 1.31$\times$10$^{-4}$ & 11 & \multirow{2}{*}{0.7416 $\pm$ 0.0002} \\
%\cline{3-8}
    & & $F = 4-3$ & 349.33899 & 41.91 & 1.28$\times$10$^{-4}$ & 9 & \\
\cline{2-8}
    & \multirow{2}{*}{$J=7/2-5/2$} & $F = 4-3$ & 349.39928 & 41.93 & 1.25$\times$10$^{-4}$ & 9 & \multirow{2}{*}{0.5611 $\pm$ 0.0002}\\
%\cline{3-8}
    & & $F = 3-2$ & 349.40067 & 41.93 & 1.20$\times$10$^{-4}$ & 7 & \\

\hline

\end{tabular}
\tablefoot{
\tablefoottext{*}{Integrated flux values and the studied lines calculated from the emission spectra in Figure \ref{imlup_emission}}
\tablefoottext{a}{Value in LAMBDA database is 340.24777 for both; new value   from \citet{Teague_2020_CN} }
\tablefoottext{b}{Value in LAMBDA database is 340.03541 for both; new value   from \citet{Teague_2020_CN}. }
}
\end{table*}

\section{Observations}

This work uses ALMA Band 7 C$_2$H ($N = 4-3$) and CN ($N = 3-2$) observations of IM Lup (Project code: \#2019.1.01357.S, PI: R.Teague) taken during the second semester of 2019 and first semester of 2021. After initial calibration through the ALMA pipeline \citep{Hunter_2023_ALMA_pipeline}, the data were flux-calibrated and self-calibrated in phase and amplitude using only the line-free spectral windows and channels with the tools of Common Astronomy Software Applications \citep[CASA; version 5.4.0,][]{McMullin_CASA} as described in the following.

The observations from each execution block (EB; seven in total) were aligned to a common phase center, considering the peak of the continuum emission of each EB as fitted by CASA task imfit. After alignment, an amplitude scaling was applied to correct for the relative flux scale variations between EBs, which were $<$10\%, following the procedures of ALMA large program DSHARP \citep{DSHARP_Andrews}. Self-calibration was then applied to the short-baseline observations (two EBs, with a longest baseline of 0.3\,km). Five rounds of phase and one round of amplitude calibration were performed, until there was no further improvement of the S/N, starting with a solution interval of 3360s and taking half the time step in each subsequent round. The S/N improvement in the short-baseline data was 726\%, from 190 to 1380. Finally, self-calibration was performed on the joint dataset of the previously self-calibrated short baselines, concatenated to the flux-calibrated long baselines (five EBs, with a maximum baseline of 1.5km). Three rounds of phase and one round of amplitude calibration were performed, the initial time solution interval was 4800s and the final S/N increase was 428\%, from 301 to 1289.

Each of the calibration solutions from the continuum analysis was then applied to the whole dataset, to correct spectral windows and channels with line emission. Using the CASA task uvcontsub, we produced continuum-subtracted measurement sets for each of our target emission lines. Each C$_2$H and CN hyperfine group was imaged using tclean with Briggs weighting and Keplerian masking\footnote{https://github.com/richteague/keplerian\_mask}. Multiple Keplerian masks, one for each of the expected hyperfine components, were used in the cleaning. A stellar mass of 1.1M$_\odot$ \citep{MAPS_Teague} and a constant $z/r$ value of 0.2 and 0.3 was used for C$_2$H and CN emission, respectively. To compromise between optimal angular resolution and high S/N, C$_2$H emission was imaged using a robust parameter of 2.0, resulting in a 0.27$\arcsec \times$ 0.25$\arcsec$ beam (PA: 64.83$^{\circ}$). CN is imaged using a 0.5 robust parameter with a final beam of 0.22$\arcsec \times$ 0.19$\arcsec$ (PA: 57.56$^{\circ}$). JvM correction \citep{MAPS_Czekala} was applied to our final images to account for the noise caused by having non-Gaussian beams. The epsilon correction factors are 0.24 and 0.52 or C$_2$H and CN, respectively \citep[for details on this procedure and the epsilon factor, refer to][]{MAPS_Czekala}. The images used in the analysis where obtained using a channel spacing of 200\,m\,s$^{-1}$ for C$_2$H and 100\,m\,s$^{-1}$ for CN, to optimize spectral resolution and S/N. The final rms of the images for each molecule are 0.6\,mJy/beam and 1\,mJy/beam for C$_2$H and CN, respectively.

Figure \ref{imlup_emission} shows the integrated intensity maps (moment 0) and stacked spectra for the data imaged at the native spectral resolution, 50\,m\,s$^{-1}$. The spectra were computed using the GoFish \citep{gofish} function, integrated\_spectrum, over the whole disk. Each of the detected hyperfine components are marked with an orange vertical line and their properties, together with the integrated fluxes indicated in Table \ref{table_molec}. A clear ring is observed in both tracers for all hyperfine groups, a detailed analysis of this feature can be found in Appendix A.

\section{Emission morphology}
\subsection{Vertical profile}

From the channel maps it is possible to visually identify the CN and C$_2$H emission originating from the upper and lower sides of the disk. Tracing the vertical location of each molecule allows us to characterize the disk structure and physical conditions of the emitting gas in defined regions. To extract the vertical profiles, we used the code ALFAHOR \citep{Paneque_2023_vert}, which traces the maxima of emission within a given masked region. The masks are visually determined and encircle the far and near sides of the disk upper surface \citep[for details on the method and terminology, see][]{Pinte_2018_method, Paneque-Carreno_2022_Elias_CN, Paneque_2023_vert}. Only channels where both near and far sides of the upper surface can be confidently identified are used for this analysis. Figures \ref{channels_cn} and \ref{channels_c2h} present the masks and location of the emission maxima for each of the occupied transitions.

\begin{figure}[h!]
   \centering
   \includegraphics[scale=0.4]{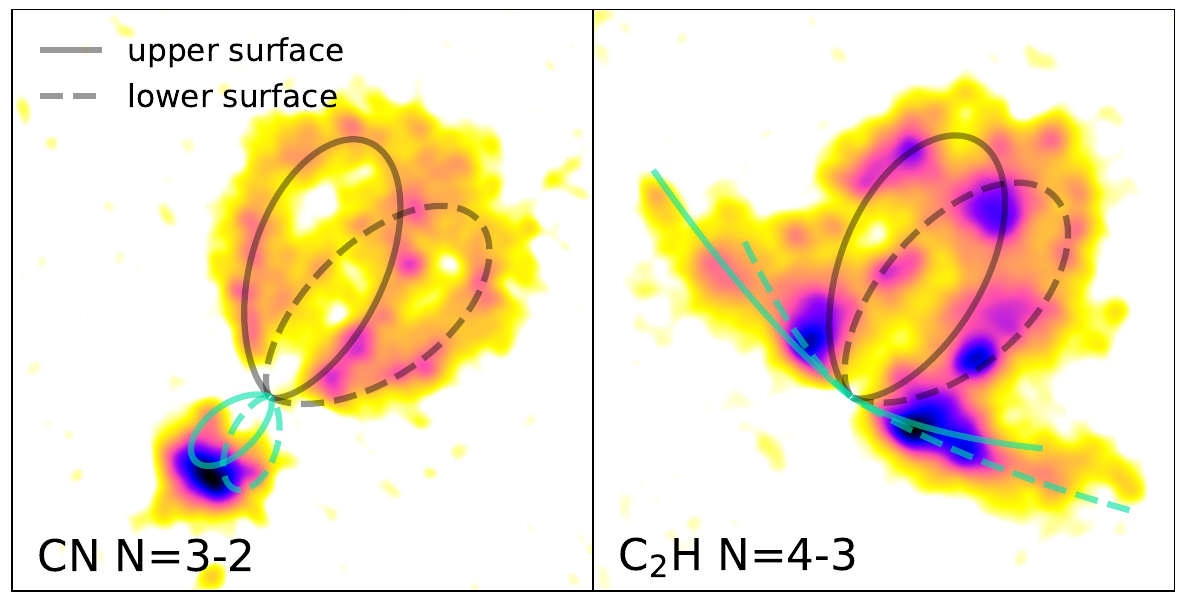}
      \caption{Channel map emission for CN (left) and C$_2$H (right) at equal relative velocity to systemic motion (-1.2\,km\,s$^{-1}$). Continuous line traces the upper emission surface and dashed line the lower surface. Different colored contours indicate different hyperfine components. Contours are not fit to the data, they represent the visual identification of the emission surfaces and follow constant $z/r$ (0.3 and 0.2 for CN and C$_2$H, respectively). 
              }
         \label{examp_height}
\end{figure}

A key assumption of this method is that the intensity maxima trace a single isovelocity along each channel. Due to the strong blending of hyperfine components in the CN emission and the low S/N of some detected transitions, we only used the unblended and strongest component CN $N=3-2 J=5/2-3/2 F= 7/2-5/2$. C$_2$H also has blending in both transitions; however, it has a sufficient separation that we can identify both components in each group (see Fig. \ref{examp_height}). However, we extracted the surface only using the $J=9/2-7/2$ hyperfine group, as it has a higher S/N. We expect that these transitions are representative of the spatial location of all the detected transitions of the same molecule, as they all have similar upper state energies (see Table \ref{table_molec}) and thus, originate from the same layer. 

\begin{figure*}[h!]
   \centering
   \includegraphics[width=\hsize]{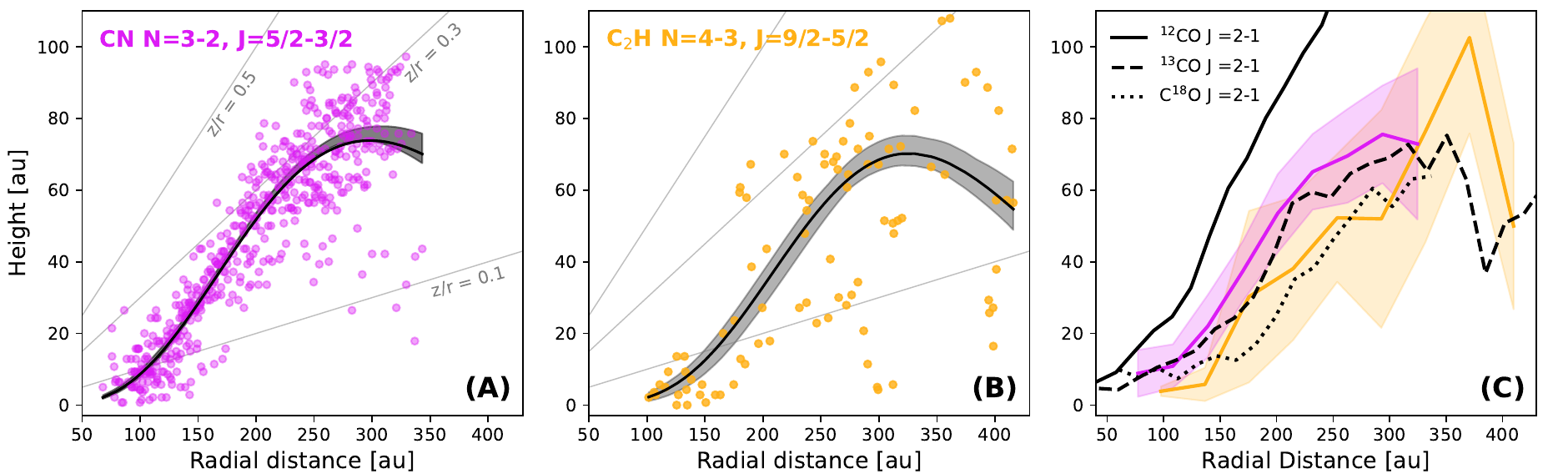}
      \caption{Vertical profiles of each tracer. Panels (A) and (B) show the data and best fit model for CN and C$_2$H, respectively (the molecule and hyperfine group used is shown in the upper left corner). Each colored dot represents the vertical location extracted at a given radius from the maxima of emission traced in the channel maps. Overlaid is an exponentially tapered model with the best-fit parameters and the statistical uncertainty of the profile. Panel (C) presents a comparison between the averaged values and standard deviation within radial bins for CN and C$_2$H (in colors) compared to the averaged CO isotopolog emission surfaces as presented in \citet{Paneque_2023_vert}. 
}
         \label{best_fit_heights}
\end{figure*}

The resulting vertical distribution is parametrized  with an exponentially tapered power law to capture the inner flared surface, plateau and turnover regions at larger radii \citep{Teague_2019_meridonial} following

\begin{equation}
z(r) = z_0\times \left(\frac{r}{100\,\mathrm{au}}\right)^{\phi} \times \exp\left[\left(\frac{-r}{r_{\mathrm{taper}}}\right)^{\psi}\right].
\end{equation}

The best-fit values for $z_0$, $\phi$, $\psi,$ and $r_{\mathrm{taper}}$ were found using a Markov chain Monte Carlo sampler as implemented by emcee \citep{emcee_ref} and shown in Table \ref{table_bestfit}. Figure \ref{best_fit_heights} shows the data considering all of the extracted maxima for CN and C$_2$H, together with the best-fit exponentially tapered power-law model.

\begin{table}[h]
\def\arraystretch{1.5}
\setlength{\tabcolsep}{5pt}
\caption{Best-fit parameters of vertical profiles.}
\label{table_bestfit}      
\centering
\begin{tabular}{c| c| c}       
\hline\hline                
  & CN $N = 3-2$ & C$_2$H $N = 4-3$ \\
\hline\hline

$z_0$ (au) & 40.29 $^{+12.36}_{-13.87}$ & 8.03 $^{+4.79}_{-3.33}$ \\
$\phi$ &  4.91 $^{+0.22}_{-0.38}$& 6.43 $^{+0.55}_{-0.84}$\\
$r_{taper}$ (au) &  65.72 $^{+26.77}_{-11.49}$ & 78.16 $^{+44.18}_{-21.86}$ \\
$\psi$ & 1.03 $^{+0.13}_{-0.06}$ & 1.18 $^{+0.23}_{-0.12}$    \\

\hline \hline

\end{tabular}
\end{table}

Comparison with previously studied CO isotopolog emission from IM Lup \citep{Paneque_2023_vert} is presented in Figure \ref{best_fit_heights} demonstrating that the location of CN ($N=3-2$) is comparable to that of $^{13}$CO ($J=2-1$). The C$_2$H ($N=4-3$) emission is located slightly below, tracing the same region as C$^{18}$O ($J=2-1$). We note that C$_2$H presents a large scatter, likely due to the lower S/N of the observations; however, our result on the average vertical profile (panel C of Fig. \ref{best_fit_heights}) is consistent with $z/r \sim$0.2 expected from the visual channel map analysis (Fig. \ref{examp_height}).

\begin{figure*}[h!]
   \centering
   \includegraphics[width=\hsize]{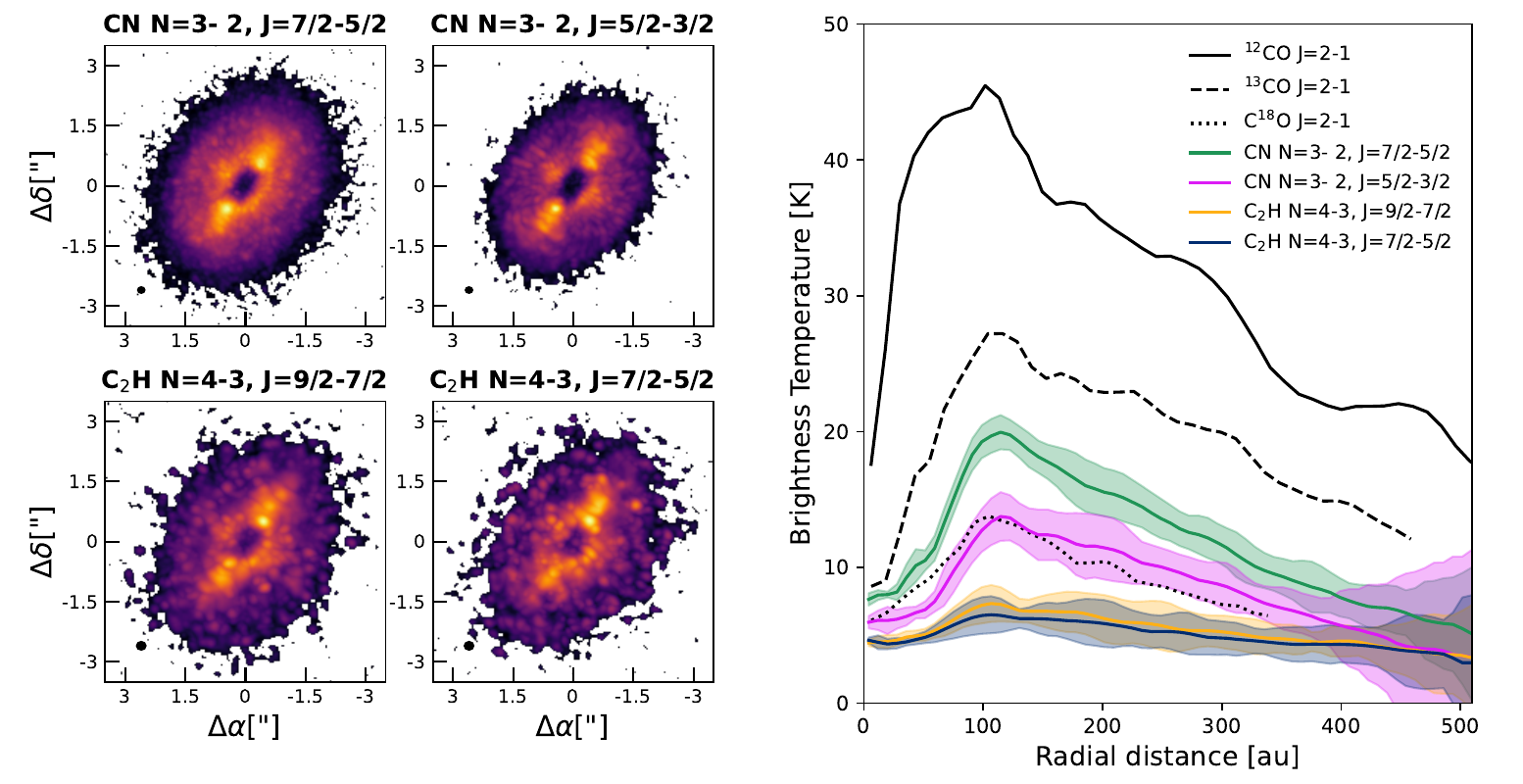}
      \caption{Brightness temperature analysis for each tracer. Left panels: Peak intensity (moment 8) maps for each hyperfine group under study. The extended feature toward the semimajor axis results from the radiative transferred effects due to higher velocity density in this region. Right panel: Brightness temperature profiles of each hyperfine group, compared to those of CO ($J=2-1$) isotopologs.}
         \label{brightness_temp}
\end{figure*}

\subsection{Brightness temperature}

Considering the best-fit vertical surfaces, the radial peak brightness temperature profiles were extracted from the peak intensity maps (moment 8). The conversion from peak intensity ($I$) to brightness temperature ($T_b$) was done using Planck's law,

\begin{equation}
    T_b = \frac{h \nu}{k} \ln^{-1}\left(1 + \frac{2h\nu^3}{I c^2}\right),
\end{equation}

\noindent where $h$ is the Planck constant, $k$ the Boltzmann constant, $c$ the speed of light, and $\nu$ the frequency of the emission. To compare the brightness temperatures of CN and C$_2$H to those of the CO isotopologs previously studied in \citet{Paneque_2023_vert}, all brightness temperature profiles were calculated using cubes with an equal spectral resolution of 0.2\,km\,s$^{-1}$ to match the resolution of the Molecules with ALMA at Planet-forming Scales (MAPS) Program data \citep{MAPS_Oberg}.

\begin{figure*}[h!]
   \centering
   \includegraphics[width=\hsize]{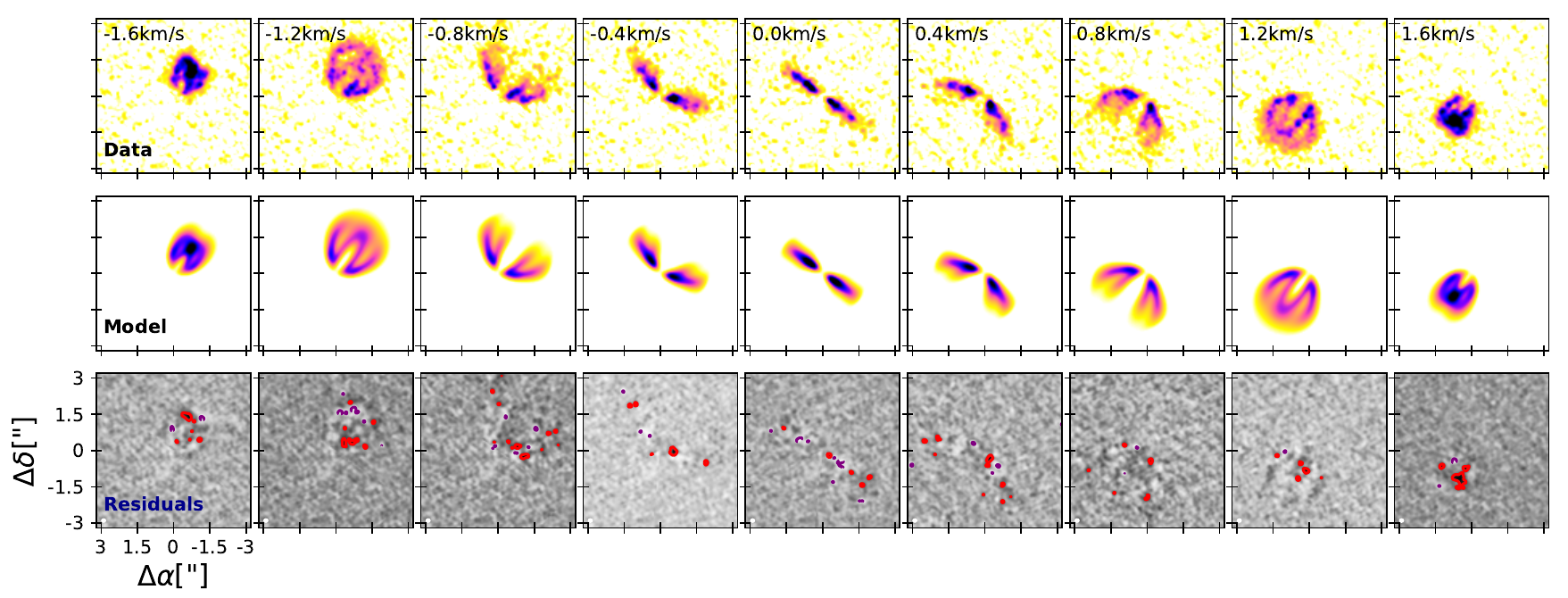}
      \caption{DISCMINER results and comparison to CN data. First row: Selection of channel maps from the CN datacube. Velocities are relative to the systemic velocity, and the spectral resolution of each channel is 0.1\,km\,$s^{-1}$. Middle row: Channel maps from the best-fit DISCMINER model, at the same spatial resolution and intensity scale as the observations. Bottom row: Residual (observations minus model) emission. Contours indicate 5$\sigma$ positive (red) and negative (purple) residuals, where $\sigma$ is the data rms (1\,mJy/beam).}
         \label{chan_discminer}
\end{figure*} 

The peak intensity maps and brightness temperature profiles, compared to those of CO isotopologs \citep{Paneque_2023_vert}, are presented in Figure \ref{brightness_temp}. The radially extended bright feature along the major axis seen in the peak intensity maps for all tracers is a radiative transfer effect related to the higher density of isovelocity contours and does not indicate a physical feature. Based on our previous analysis we know that CN and C$_2$H emission is spatially colocated with $^{13}$CO; however, both transitions show lower brightness temperatures than $^{13}$CO for all hyperfine groups. The drop in temperature at the innermost radii is dominated by beam dilution as the emitting area becomes comparable to or smaller than the spatial resolution; however, this drop extends out to $\sim$150\,au, which is beyond the beam size. Dust absorption is likely a big contributor to the low temperatures, as several works have shown that the inner disk of IM Lup is optically thick \citep{Cleeves_LupusModel_2016, sierra_2021_MAPS}.

The brightness temperature may be a tracer of the excitation temperature of the region, but only in the cases where the emission is optically thick ($\tau>>$1), as it is proportional to $T_{ex}(1 - e^{-\tau})$. CO is expected to be abundant in IM Lup, in particular $^{12}$CO and $^{13}$CO should be optically thick \citep{Cleeves_LupusModel_2016} and therefore their brightness temperature may be taken as a tracer of the kinetic temperature in the region. On the other hand, CN and C$_2$H are expected to have lower column densities and be optically thin tracers \citep{Cazzoletti_2018_CN, MAPS_Bergner, Bergin_2016_c2h}, which is corroborated by the lower brightness temperatures measured in this work. We note that the temperature difference between both CN hyperfine groups is due to the strong blending in the $J=7/2-5/2$ transition (see Fig. \ref{imlup_emission}), which means that the peak flux for a given pixel is the sum of multiple hyperfine components. This is the same situation that affects the integrated flux values shown in Table \ref{table_molec}.

\section{Parametric emission model: DISCMINER fit}

Knowing that the molecular emission is optically thin enables us to directly determine turbulence from the nonthermal contributions to the linewidth \citep{Hacar_2016_optdepth}. This is not possible in optically thick tracers (such as CO isotopologs) due to the additional opacity broadening that affects the linewidth in those cases, which is why in these cases radiative transfer models are needed to account for the opacity effect. Extracting the linewidth is not straightforward, it requires a parametric model that accounts for the vertically elevated emission, the upper and lower sides as well as radial intensity and velocity gradients. To this end, we used DISCMINER \citep{izquierdo_2021_discminer1}, which allowed us to find the best-fit parametrization of our emission.

\begin{figure*}[h!]
   \centering
   \includegraphics[width=\hsize]{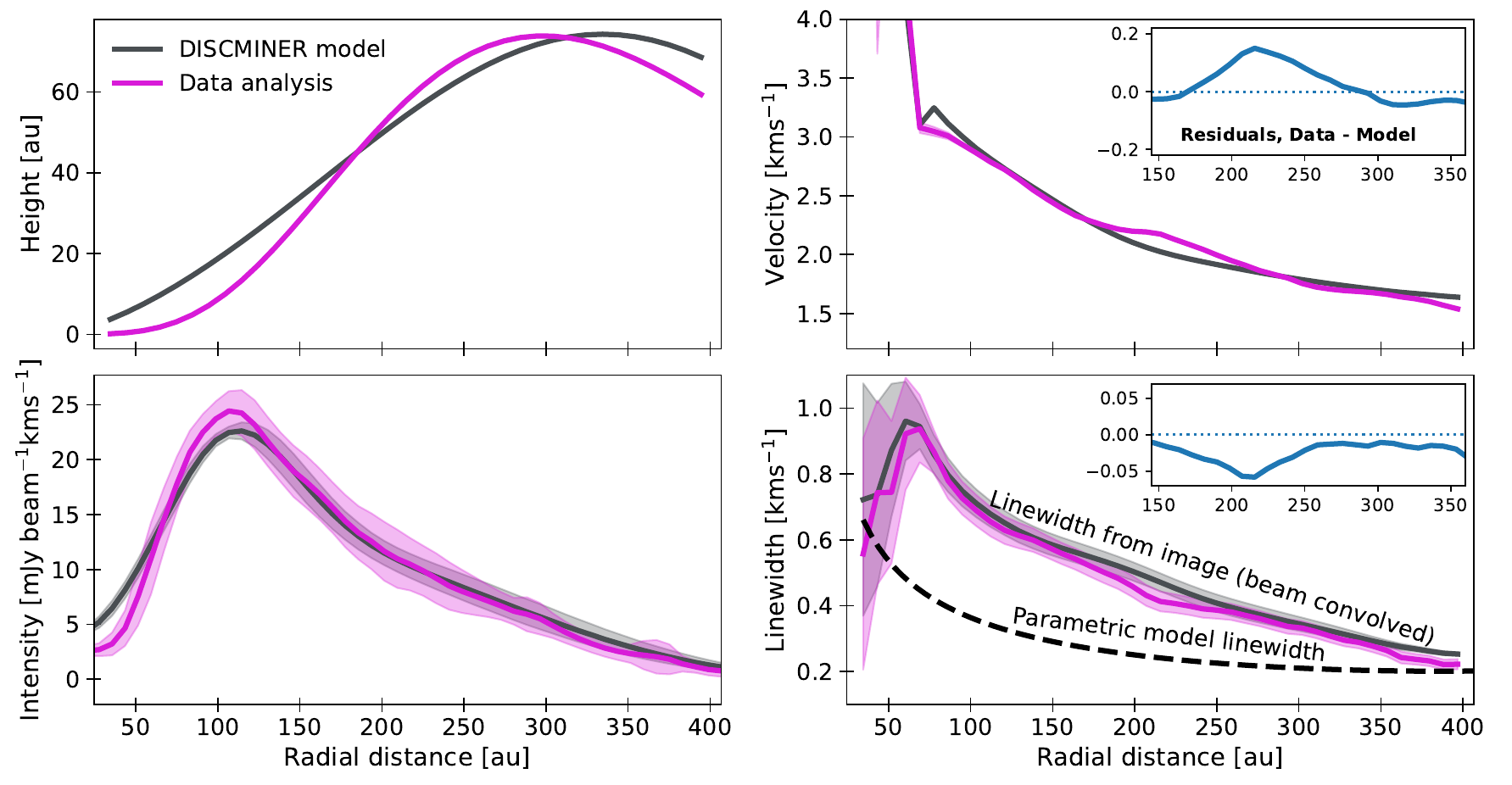}
      \caption{Comparison between various radial profiles between the data (magenta curves) and the DISCMINER model (gray curves). All panels except the height profile indicate the profile dispersion as a shaded region. Velocity and linewidth panels have a noticeable difference located at $\sim$220\,au, a blue residual curve characterizes this feature in the upper right corner of each panel. The linewidth panel shows the measured linewidth extracted from the data and the beam-convolved model by fitting the spectra. Additionally, the dashed black line indicates the intrinsic parametric form of the linewidth, as fitted by DISCMINER. The beam-convolved model will always have a broader linewidth than the parametric form and is used to directly compare our model to the data, which has a finite sptial resolution. However, the intrinsic linewidth is what we used in our calculations, as this is expected to be the actual line broadening due to thermal and nonthermal motions.}
         \label{comp_discminer_obs}
\end{figure*}

\subsection{Model setup}

DISCMINER determines the best orientation, velocity, surface and line profile parameters for the observed emission by creating a model intensity, $I_m$ as a function of the disk cylindrical coordinates ($R, z$) and the velocity. The emission kernel has two components, one for the upper and another for the lower surface, which are well separated in the case of our studied emission. Contrary to previous works with DISCMINER, which focus on CO isotopolog emission \citep{izquierdo_2021_discminer1, Izquierdo_2021_hd16planet, Izquierdo_2023_discminer2}, this work traces optically thin molecules that will be well reproduced by a Gaussian line profile \citep{Hacar_2016_optdepth}. Therefore, we used a Gaussian kernel for each surface such that

\begin{equation}
  I_m(R,z;v_{chan}) = I_p\, \mathrm{exp}\left[-\frac{(v-v_{chan})^2}{\Delta V^{2}}\right],
\end{equation}
 
\noindent with $I_p$ being the peak intensity and $\Delta V$ the linewidth at a given radial and vertical distance from the star. Since we assumed fully optically thin emission, the final model intensity is given by the direct sum of the upper and lower disk surface intensity contributions and compared against the emission spectra extracted from each pixel of our observational data cube \citep[for specific details, see][]{izquierdo_2021_discminer1, Izquierdo_2023_discminer2}. Table \ref{table_discminer} indicates the equations used to model each of the disk attributes. Each emission surface is modeled independently as an exponentially tapered power law, the intensity is considered to be a power law with a Gaussian component, to account for the ring in the integrated emission maps (see Figure \ref{imlup_emission}). The linewidth is assumed to follow a simple radial and vertical power law (as indicated in Table \ref{table_discminer}).

\begin{table}[h]
\def\arraystretch{2.2}
\setlength{\tabcolsep}{6pt}
\caption{Parametric model attributes.}
\label{table_discminer}      
\centering
\begin{tabular}{l c}       
\hline\hline                
  Attribute & Prescription \\
\hline\hline

Orientation & $i$, PA, $x_c$, $y_c$\\
\hline
Velocity & $v_{\mathrm{rot}}=\sqrt{\frac{GM_*}{r^3}}R$, $v_{\mathrm{sys}}$\\
\hline
Upper Surface & $z_{\mathrm{up}} = z_0 (R/D_0)^p $ exp$[-(R/R_t)^q]$\\
Lower Surface & $z_{\mathrm{low}} = -z_0 (R/D_0)^p $ exp$[-(R/R_t)^q]$\\
\hline
Peak intensity & $I_p = I_0 (R/D_0)^{p} (z/D_0)^{q} + I_g$ exp $\left[-\frac{(R - R_g)^2}{2 \sigma_{g}^{2}}\right ]$\\
\hline
Line width & $\Delta V = L_0(R/D_0)^{p} (z/D_0)^{q}$\\

\hline \hline               
\end{tabular}
\tablefoot{$G$ is the gravitational constant and $D_0$ = 100 au is a normalization factor. $R$ is the cylindrical radius and $z$ is the height from the midplane, always positive, $r$ is the spherical radius. The other variables are free parameters, even if named alike they are modeled independently for each attribute.}                     

\end{table}

DISCMINER is not built to fit emission with multiple hyperfine components; therefore, we created a cube that only covers the emission of the strongest unblended hyperfine component of CN 3-2, J = 5/2 - 3/2 F= 7/2-5/2 (see Figure \ref{imlup_emission}). It is not possible to analyze the other CN hyperfine components, due to the strong blending of the high S/N lines, and the low signal of the other detections. 
C$_2$H is also excluded from this analysis due to its low S/N and the close blending of the lines, which does not allow us to isolate them as we can for CN (see Figure 3 for a comparison on how close the CN hyperfine components are compared to C$_2$H in the channel map emission). Future work incorporating hyperfine components to the DISCMINER kernel could allow us to optimize our model using the whole dataset, but it is beyond the scope of this study. 

\subsection{Comparison to CN observations}

The best-fit model parameters are found in Appendix C. Figure \ref{chan_discminer} shows the resulting model channel maps, convolved with a Gaussian beam to be at the same spatial resolution as the observations. The 5$\sigma$ residuals are presented in the bottom row, no coherent or systematic pattern is identified. Figure \ref{comp_discminer_obs} shows further comparison between the model and data. There is good agreement between the derived vertical and azimuthally averaged intensity profiles and the DISCMINER parametric forms. The rotation curve shows a $\sim$200m\,s$^{-1}$ residual between 200 and 250\,au, which coincides with a  $\sim$50m\,s$^{-1}$ negative residual in the linewidth. This velocity variation is also observed in $^{12}$CO \citep[see Figure B.18 in][]{Izquierdo_2023_discminer2} and the location of the feature is close to a dust gap at 209\,au and a dust ring at 220\,au \citep{DSHARP_Huang_radial}, which may be related to these variations.

We note that the linewidth values extracted from the channel map images (red and gray curves, with uncertainty regions in lower right panel) are higher than the best-fit DISCMINER parametric form (traced by dashed black line). This is because the profiles extracted from the image correspond to direct measurements of the linewidth from the spectra extracted at each pixel in the image plane. These measurements will always be broader (have higher values) than the underlying molecular emission linewidth due to the linewidth broadening caused by the finite spatial resolution of the observations \citep[usually called beam broadening and discussed also in][]{Teague_2020_CN}. When we refer to the DISCMINER parametric form, we are describing what is expected to be the native molecular linewidth that, convolved with a Gaussian beam equal to the spatial resolution of the ALMA images ($\sim$0.22$\arcsec$), will match the broader values extracted directly from the data. 

\section{Turbulence as traced by CN}

The measured linewidth of an optically thin tracer will be dominated by doppler broadening caused by thermal and nonthermal motions \citep{Hacar_2016_optdepth}. As we used a Gaussian kernel (see Eq. 3), our measured linewidth $\Delta V$ corresponds to the velocity dispersion such that the full width at half maximum (FWHM) of the kernel is $2\sqrt{ \mathrm{log}2} \,\Delta V$. The linewidth can then be written as

\begin{equation}
    \Delta V^2 = V_{\mathrm{th}}^2 + V_{\mathrm{non-th}}^2.
\end{equation}

The thermal component is well characterized and proportional to the ratio between the kinetic temperature and the molecular mass of the observed tracer \citep{Myers_Ladd_Fuller_1991}. The nonthermal component can originate from non-Keplerian motion such as photo-evaporative winds, planet--disk interactions, or turbulence \citep[e.g.,][]{Fuller_Myers_1992, Hacar_2016_optdepth, Izquierdo_2023_discminer2}. For this analysis we considered the case where nonthermal motion is fully due to turbulence; the implications and possibilities of other origins are reviewed in the discussion section. Therefore, the two components of the measured linewidth can be decomposed as

\begin{equation}
    \Delta V^2 =  \frac{2 k T_{kin}}{m_X}  + \varv_{\mathrm{turb}}^{2} .
\end{equation}

\noindent Here $k$ is the Boltzmann constant and $m_X$ the molecular mass, in this case, for CN, $m_X$ = 26.0 \citep{Klisch_1995_CN, lambda_database_Schoier_2005}. The nonthermal broadening is assumed to be completely due to turbulence, $\varv_{\mathrm{turb}}$. Turbulence can then be expressed in terms of the Mach number, $\mathcal{M}$, such that

\begin{equation}
    \mathcal{M} =  \frac{\varv_{\mathrm{turb}}}{\sqrt{2} c_s}
,\end{equation}

\noindent where $c_s$ is the local sound speed in the studied layer $c_s^2 = \frac{k T_{kin}}{\mu m_H}$ \citep{Balbus_Hawley_1998, Flaherty_2015_hd16_weakturb}. We generally refer to turbulence in terms of the Mach number so it may be directly compared to previous works and estimations \citep{Flaherty_2015_hd16_weakturb, Flaherty_2017_DCOp, Flaherty_2020_DMTau, Teague_2016_turb_TWHya, Teague_2018_turb_nonlte}.

\begin{figure}[h!]
   \centering
   \includegraphics[width=\hsize]{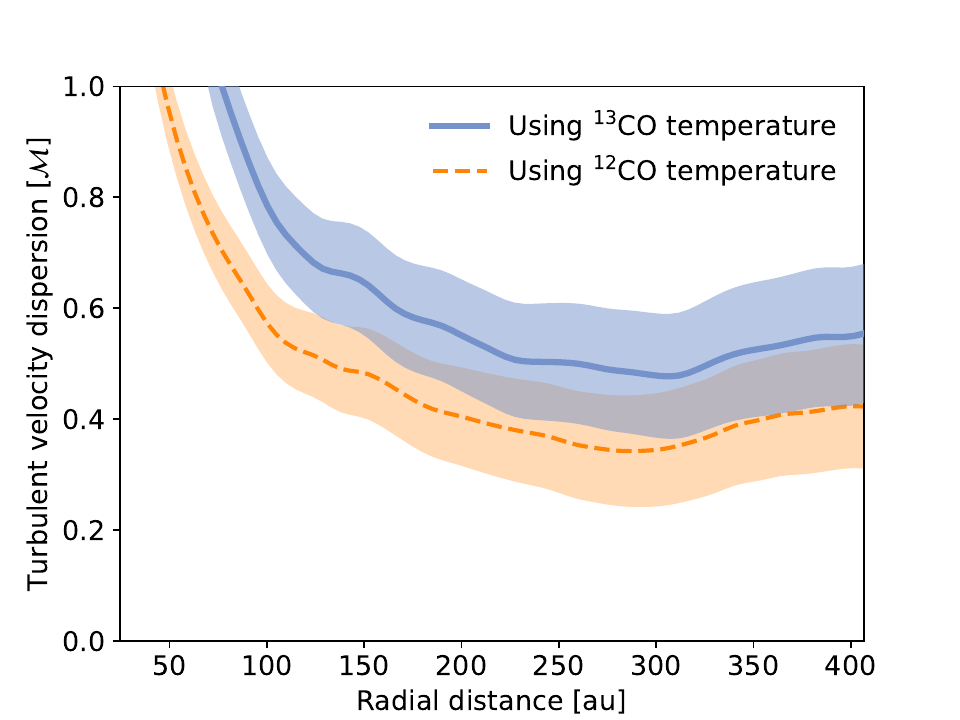}
      \caption{Turbulence extracted from the nonthermal component of the CN ($N=3-2$) linewidth. The blue curve indicates results when considering the brightness temperature of $^{13}$CO. The orange curve traces results using the brightness temperature of $^{12}$CO. Shaded regions represent the dispersion when considering an error of 50\,m\,s$^{-1}$ on the measured linewidth.
              }
         \label{turb_imlup}
\end{figure}

By spatially localizing the CN emission vertically and radially close to the optically thick CO isotopes, we have a tool to directly access the kinetic temperature without need to use parametric models or additional molecular transitions. Using the linewidth parametrization found by DISCMINER we can immediately calculate the turbulence radial profile, presented in Figure  \ref{turb_imlup}. Our results indicate turbulent motions of $\mathcal{M}=$0.4-0.6 at radii beyond $\sim$150\,au. The shaded regions of the plot represent the variation of our results considering a 50\,m\,s$^{-1}$ error in the linewidth measurement. While the statistical error of our fit is $<$5\,m\,s$^{-1}$, we took the native resolution of our data as the largest possible deviation, in case effects such as Hanning smoothing are significantly affecting  our linewidth extraction \citep{ALMA_tech_C10}. Under the $\alpha$-prescription for turbulent viscosity \citep[$\nu = \alpha c_s H$,][]{Shakura_Sunyaev_1973_visc} it is possible to relate the turbulence Mach number $\mathcal{M}$ to $\alpha$ following, $\alpha \sim \mathcal{M}^{2}$ \citep{Rosotti_2023_turbulencerev}. Our measurements indicate $\alpha \sim 10^{-1}$.

The total measured linewidth ($\Delta V$) and turbulence values increase steeply in the inner $\sim$150\,au. While IM Lup has been proposed to be extremely turbulent in the inner disk \citep{Cleeves_LupusModel_2016, Bosman_2023_imlup} it is unclear if our analysis is sensitive to this effect, or if the high turbulence should extend beyond the innermost 20\,au, given the lower estimated scale height for millimeter dust in this system \citep{Bosman_2023_imlup}. Linewidth measurements are strongly affected by overlapping velocities in the inner disk \citep{Teague_2020_CN, Izquierdo_2021_hd16planet}; therefore, we only considered the values beyond 150\,au as accurate measurements and do not refer to the high values inward of 150\,au. 

\begin{figure}[h!]
   \centering
   \includegraphics[width=\hsize]{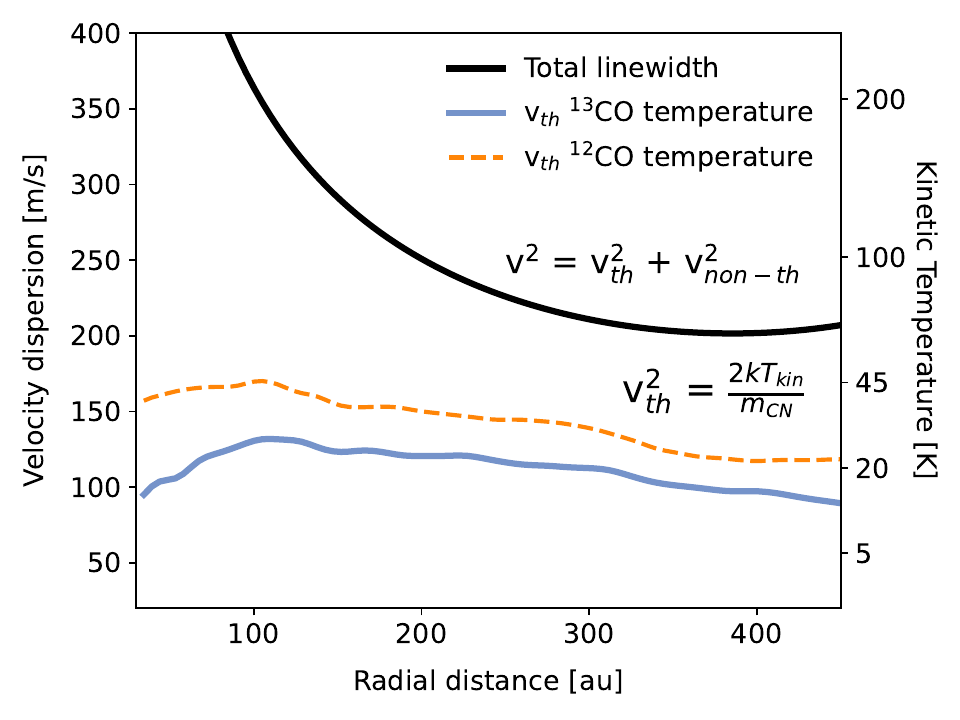}
      \caption{Total intrinsic linewidth determined from the best-fit DISCMINER model (solid black line; this is the same as the dashed black line in the bottom-right panel of Figure \ref{comp_discminer_obs}).\ It is compared to the thermal component obtained using the $^{13}$CO (blue) or $^{12}$CO (orange) brightness temperatures. 
              }
         \label{temp_linewidth}
\end{figure}

\begin{figure*}[h!]
   \centering
   \includegraphics[width=\hsize]{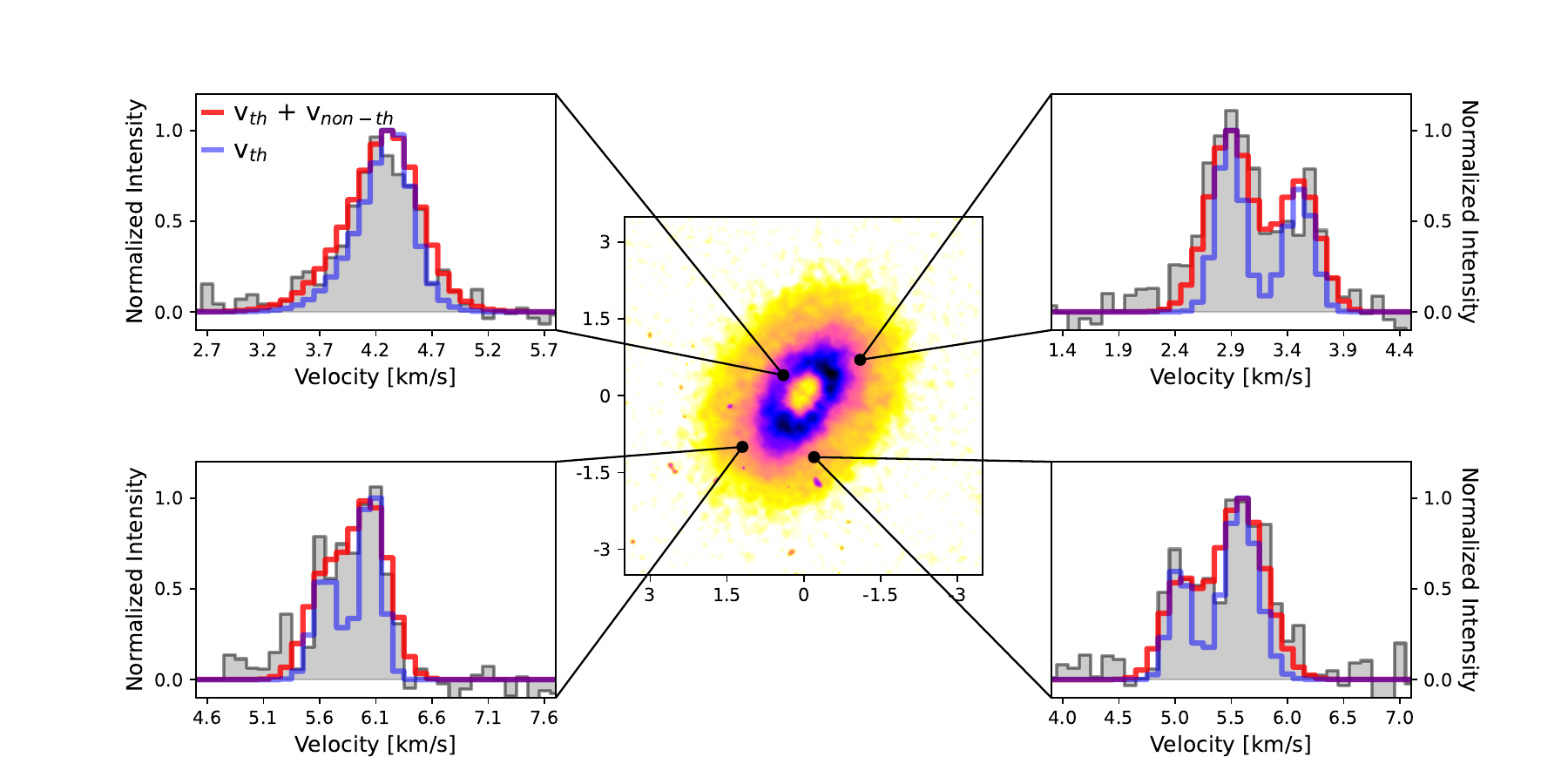}
      \caption{Spectra of four different pixels in the CN disk emission (in gray), indicated on the central integrated moment map. Overlaid are the DISCMINER model spectra from the best-fit parameter fit, considering the total measured linewidth (thermal and nonthermal components) in red. For comparison, the best-fit DISCMINER model considering a linewidth from only the thermal component calculated using the $^{13}$CO temperature is overlaid* in blue.
              }
         \label{spectra_pix_rich}
\end{figure*}

The average total linewidth measured outward of 150\,au is $\sim$221 m\,s$^{-1}$. Thermal broadening only accounts for 105-132\,m\,s$^{-1}$, using the $^{13}$CO and $^{12}$CO brightness temperatures, respectively. Therefore, the nonthermal broadening ranges between 177 and 194\,m\,s$^{-1}$ and would require temperatures $\sim$40\,K higher than measured to account for it as thermal motion. Figures \ref{temp_linewidth} and \ref{spectra_pix_rich} illustrate the need of a high nonthermal component in order to explain the measured total linewidth (see Fig. \ref{temp_linewidth}) and the data spectra (see Fig. \ref{spectra_pix_rich}).

\section{Discussion}

\subsection{Vertical location of CN and C$_2$H}

Our work traces the CN and C$_2$H emission originating from a vertical region similar to $^{13}$CO, at $z/r=$0.2-0.3 constrained radially between 50 and 350\,au.  These results are in agreement with chemical models of IM Lup that locate the tracer at $z/r = $0.2 \citep{Cleeves_2018}. However, while in agreement with theoretical models, this is the first direct observational evidence of C$_2$H coming from vertically elevated layer in the outer disk, as previous works studying other protoplanetary disks had only traced C$_2$H inward of 150\,au and closer to the midplane \citep{MAPS_Law_Surf, Paneque_2023_vert}. 

Both CN and C$_2$H are formed via UV-driven chemistry \citep{Nagy_2015_C2H, Bergin_2016_c2h, Visser_2018, Cazzoletti_2018_CN, Miotello_2019_c2h, MAPS_Guzman, MAPS_Bergner} and tracing their vertical distribution in various transitions will allow us to further understand the range of penetration that these energetic photons have toward the midplane, or the mixing processes that may result in some transitions tracing different vertical zones. The distribution of the small grains may have a particularly strong effect on how the UV field affects the disk \citep{Bergin_2016_c2h, maps_bosman}. Combining scattered light observations with high resolution ALMA data of UV-sensitive molecular lines will give insight into these effects. Overall, future work should focus on characterizing differences in the vertical location of various transitions for a single system, particularly of CN and C$_2$H. IM Lup is the perfect laboratory for these studies.

\subsection{Origin of high turbulence in IM Lup}

The physical structure of IM Lup has been thoroughly studied with models \citep[e.g.,][]{Cleeves_LupusModel_2016} that propose the existence of a turbulent inner disk \citep[<20\,au][]{Cleeves_2018, Bosman_2023_imlup} with high dust optical depth \citep{DSHARP_Huang_radial, sierra_2021_MAPS} that may be causing the cavity seen in molecular emission. Additionally, a series of instabilities may be ongoing in IM Lup. \citet{Bosman_2023_imlup} proposed that the turbulent inner disk could be related to vertical shear instability \citep{Urpin_Brandenburg_1998_VSI}. The high disk-to-star mass ratio measured in multiple works \citep[0.1-0.2, ][]{Lodato_2023_mass_imlup, Cleeves_2018, MAPS_Zhang, sierra_2021_MAPS} suggests gravitational instabilities \citep[][]{Kratter_Lodato_2016_GI} throughout some portion of the system. Finally, analysis of the ionization structure suggests a radial gradient in the cosmic ray ionization rate and adequate conditions for magneto-rotational instabilities \citep[MRIs;][]{Balbus_1991_MRI} in disk regions beyond the millimeter continuum disk edge ($\sim$313\,au) or above 20\,au or $z/r \sim$ 0.25 from the midplane \citep{Seifert_2021_MRI_imlup}.

Indeed, high turbulence in this system is not surprising. Any of these instabilities, or their combined effect, could be causing turbulent motions. There have already been suggestions of nonthermal broadening from low-resolution CO data, as thermochemical models are able to explain the CO morphology using 100\,m\,s$^{-1}$ turbulent velocities \citep{Cleeves_LupusModel_2016} a value within the order of our derived range of 177-194\,m\,s$^{-1}$. In contrast with this, low turbulence values ($\alpha \sim 10^{-3}$) are  derived from simultaneous modeling of millimeter (ALMA) and micrometer (SPHERE) dust observations \citep{Franceschi_2023_imlup_turbdust, Jiang_turb_dust}. This low turbulence is mainly driven by the millimeter dust distribution, which traces the cold midplane. Through our analysis, we constrain high turbulence values ($\alpha \sim 10^{-1}$) traced by CN at $R>$150\,au and in the vertically elevated disk layers (above the midplane) at $z/r = 0.25-0.3$, which would suggest a vertical gradient in the turbulence, with higher values in the upper disk and lower values toward the midplane. Theoretical MRI studies have shown that high $\alpha$ values may correlate with higher accretion rates \citep{Simon_2015}. IM Lup has a measured disk accretion rate of 10$^{-8}$ M$_{\odot}$ yr$^{-1}$ \citep{alcala_2017_accretion}, which is typical of lower $\alpha$. This is reminiscent of DM Tau, where the system showed measurable turbulence, but a low accretion rate \citep{Flaherty_2020_DMTau}. It is important to note that the measured accretion rate is taken from the instantaneous accretion rate onto the star; however, the $\alpha$ values represent disk motions throughout the outer, and in the case of IM Lup the elevated, regions of the disk. Future work should look into this with dedicated models.

A vertical gradient in the turbulence, with high values approaching the sound speed above 2$H$, with $H$ being the disk hydrodynamical scale height, is a key prediction of MRIs \citep{Simon_2015, Flock_2015}. The scale height in IM Lup is expected to be $H \sim 0.1$ \citep{MAPS_Zhang, Paneque_2023_vert}; therefore, CN emission is originating above 2$H$. To fully confirm a vertical gradient in turbulence, studying molecules in the intermediate layers is necessary, such as HCO$^{+}$, H$_2$CO, and HCN, which have their vertical heights characterized and emit from in-between the midplane and $^{13}$CO \citep{Paneque_2023_vert}. Unfortunately, the sensitivity and spectral resolution of the available data for these tracers \citep[$\leq$200m\,s$^{-1}$, ][]{MAPS_Oberg} is not sufficient to detect nonthermal broadening of $\leq$90m\,s$^{-1}$, such as detected for CN. Follow up observations at higher spectral resolution would allow us to replicate this analysis and confirm the existence of a vertical gradient. 

\begin{figure}[h!]
   \centering
   \includegraphics[width=\hsize]{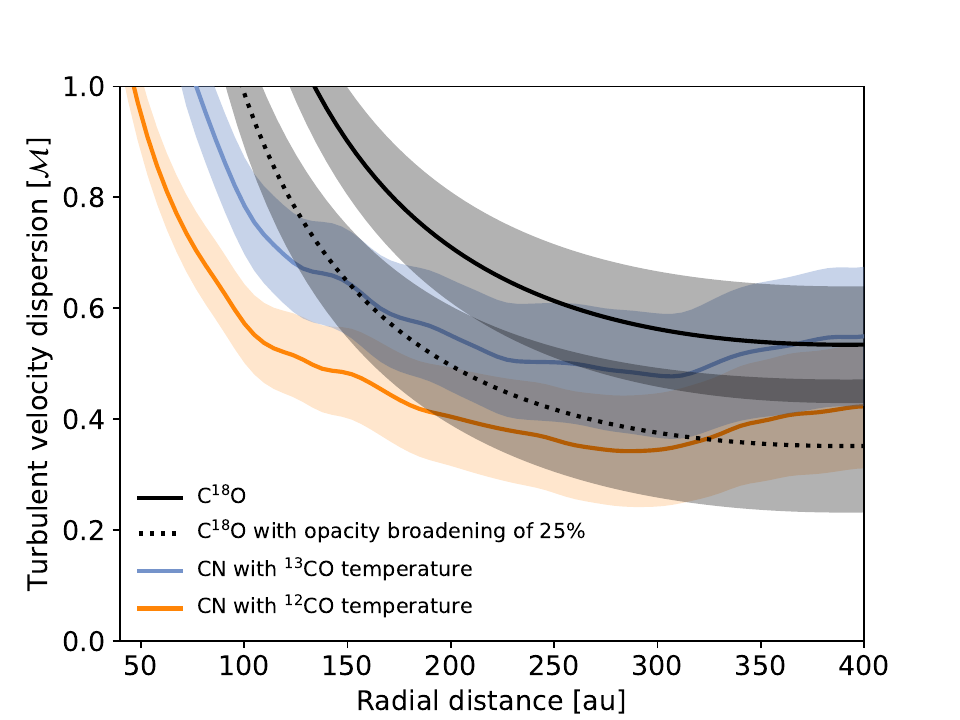}
      \caption{Same as Figure 7, but including an analysis of the nonthermal component of C$^{18}$O $J=2-1$ data. The solid black line traces the results when applying the same methodology as previously used in CN to C$^{18}$O. The dashed black line is the turbulence profile obtained after assuming a 25\% opacity broadening in the measured linewidth, which may be the case if C$^{18}$O has optical depths of 1-2.
              }
         \label{turb_C18O}
\end{figure}

As a first attempt to compare our results with other less abundant tracers, we repeated our analysis on C$^{18}$O $J=2-1$ data from the MAPS program \citep{MAPS_Oberg}. Figure \ref{best_fit_heights} shows that C$^{18}$O $J=2-1$ comes from an elevated layer, comparable to that of CN, but slightly below. This emission is expected to come close to the CO freeze-out region and previous works have assumed an excitation temperature of 21\,K for the C$^{18}$O gas \citep{Pinte_2018_method, Paneque-Carreno_2022_Elias_CN}. Assuming that the nonthermal contribution is fully turbulent, the extracted turbulent dispersion profile of C$^{18}$O are slightly higher, but comparable within the uncertainty to the results obtained from CN assuming the temperature profile of $^{13}$CO. We note that while the brightness temperature of C$^{18}$O is below the assumed excitation temperature (see Figure \ref{brightness_temp}), the column density of CO in IM Lup \citep{MAPS_Zhang} and estimates from Elias 2-27 of C$^{18}$O under similar conditions \citep{Paneque-Carreno_2022_Elias_CN}, show that this isotopolog may have optical depths of 1-2. In this case, opacity broadening could account for up to 25\% of the measured linewidth value \citep{Hacar_2016_optdepth}, biasing our result to higher turbulence profiles. Figure \ref{turb_C18O} shows the comparison between the CN results and the turbulent dispersion profiles of C$^{18}$O assuming a fully turbulent nonthermal contribution in the same way as our CN analysis (solid line) and considering an opacity broadening of 25\% to the measured linewidth (dotted line). While the analysis of C$^{18}$O considering opacity broadening gives a turbulent dispersion profile below that of CN (using the $^{13}$CO temperature), all profiles are comparable within their uncertainty. This is likely due to the similarity in the emitting region, which overlap within the dispersion \citep[for details, see][]{Paneque_2023_vert} and highlights the need for a follow up study using other optically thin tracers that clearly probe different innermost disk layers.

\subsection{Possible overestimation of the nonthermal component}

We considered the possibility that we overestimated the turbulence in our analysis, for example by underestimating the temperatures in the CN emitting region or not taking into account alternative origins of the nonthermal broadening. Regarding the temperature profiles, we are confident that CO traces optically thick gas, even toward the outer disk regions. We confirmed this by comparing our temperature profiles to the gas temperature from dedicated thermochemical models of IM Lup \citep[see Figs. 8 and 9 of][]{Cleeves_LupusModel_2016}, which coincide with the values used in this work; therefore, we have not underestimated the temperature. Determining other contributors to nonthermal broadening is harder: it could be that CN emission is not as optically thin as expected, that it is vertically extended and the velocity dispersion along the line of sight is considerable, or that some strong non-Keplerian motion is affecting the system. Below, we analyze each of these scenarios.

At optical depths of $\tau\sim$0.1 the opacity contribution to the linewidth is negligible, but at slightly higher values, even while $\tau<$1 as expected for CN in IM Lup \citep{MAPS_Bergner}, opacity broadening could contribute up to $\sim$15\% of the measured linewidth \citep{Hacar_2016_optdepth}. For our measured linewidth of $\sim$221\,m\,s$^{-1}$ outside of 150\,au we considered a systematic error of 50\,m\,s$^{-1}$ to account for Hanning smoothing effects, which corresponds to a 23\% error in the linewidth measurement and also accounts for the possibility of having CN emission with an optical depth up to $\tau\sim$2. Our results would still be consistent with high turbulence, even if the emission were marginally optically thick and opacity was contributing to the nonthermal broadening. 

CN emission is expected to arise from a geometrically thin layer \citep{Cazzoletti_2018_CN, Paneque-Carreno_2022_Elias_CN}; however, at the peak of emission this layer may extend vertically up to $\sim$40\,au \citep[see Figure 3 from][]{Cazzoletti_2018_CN}. It is not possible to estimate the vertical width of the CN emission with our current analysis, but we can test the impact on our measured linewidth caused by Keplerian shear along the line of sight for IM Lup. By considering the inclination of the system and taking the best-fit CN emission layer as the position of the lowest parcel of material, we measured the linewidth broadening caused by having different vertical widths. The detail of this analysis can be found in Appendix D. Figure \ref{width_textsingle} comprises the results, showing that at radial distances beyond 150\,au the variation in the linewidth caused by vertical widths of up to 50\,au is always $\leq$25\,m\,s$^{-1}$. The width would have to be broader than $\sim$60\,au to cause significant variations out to further radii, which is not expected for CN \citep{Cazzoletti_2018_CN}. This indicates that even in the case of CN being vertically extended, our results beyond 150\,au will not vary beyond our assumed uncertainty of 50\,m\,s$^{-1}$ (Fig. \ref{turb_imlup}).

\begin{figure}[h!]
   \centering
   \includegraphics[width=\hsize]{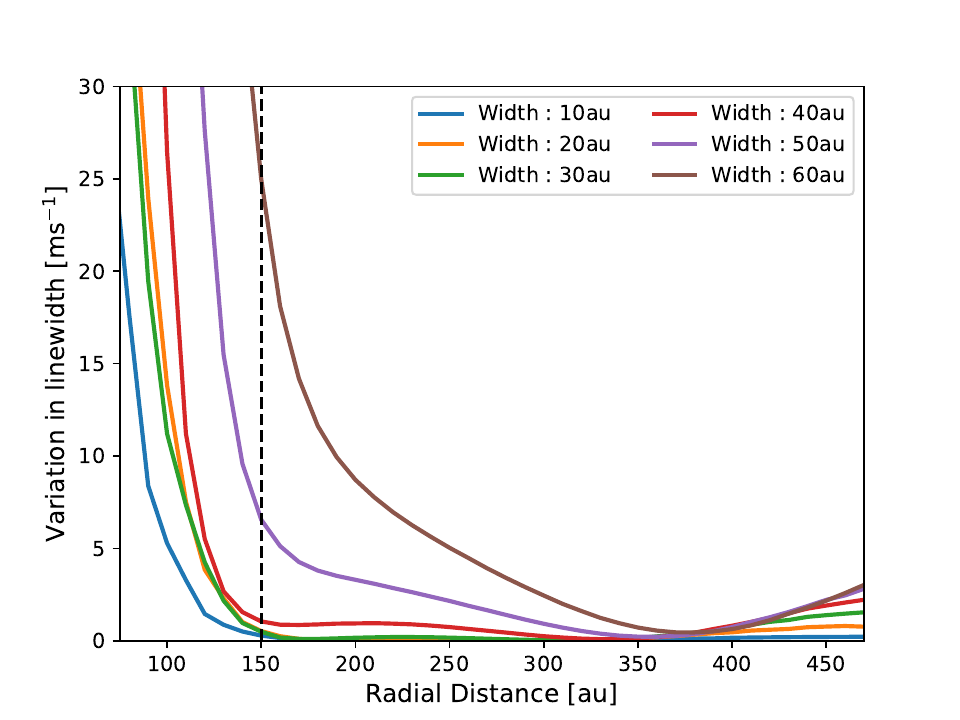}
      \caption{Linewidth broadening due to the vertical Keplerian shear, for different vertical widths. Each colored line indicates a specific emission width. The vertical dashed line at 150\,au indicates the radial distance beyond which we consider our nonthermal linewidth component measurements robust. Details on the method are found in Appendix D.
              }
         \label{width_textsingle}
\end{figure}

There are no immediate signs of non-Keplerian motion \citep[as seen in other systems, for example HD\,163296;][]{Pinte_2018_hd16planet, Teague_2018_hd16planet, Izquierdo_2021_hd16planet} beyond the strong residual at $\sim$220\,au (Figure \ref{comp_discminer_obs}), which is in any case radially localized and would not change our main results across the complete radial extent. However, IM Lup has been found to be prone to strong photo-evaporative winds, even in the presence of a weak external radiation field. Theoretical work by \citet{Haworth_2017_imlupwind} shows that, if present, these winds would originate in the upper disk layers and possibly cause mass loss rates comparable to the disk accretion rate \citep[10$^{-8}$ M$_{\odot}$ yr$^{-1}$,][]{alcala_2017_accretion}, due to the large vertical extent of the disk, which makes the surface material weakly gravitationally bound. Winds are expected to broaden the line \citep{Fuller_Myers_1992}; however, as we do not detect any indication of them in the spectra or image plane, we did not include this uncertainty in our analysis.

\subsection{Comparison with other methods and results}

We have presented a new methodology to directly extract turbulence, a fundamental disk property, from molecular line emission. By tracing the vertical location of an optically thin tracer and comparing it to that of optically thick CO lines we can immediately access the kinetic temperature of the region \citep{MAPS_Law_Surf, Law_2023_12CO_surf, Paneque-Carreno_2022_Elias_CN} and from the measured linewidth calculate the nonthermal contribution attributed to turbulent motions. This requires a precise parametric model of the spectra to extract the linewidth accounting for the upper and lower disk emission layers and the broadening caused by beam convolution, which is why we use DISCMINER \citep{izquierdo_2021_discminer1}.

Previous works have mostly put upper limits on turbulence by modeling the emission using physical disk models \citep{Hughes_2009_turb, Guilloteau_2012_turb_dmtau, Flaherty_2015_hd16_weakturb, Flaherty_2017_DCOp, Flaherty_2020_DMTau}. This approach is required when the observed tracer is optically thick, as the linewidth will be affected by an opacity broadening due to the column density of the molecule \citep{Hacar_2016_optdepth} and the comparison between model and data must be done taking into consideration radiative transfer calculations \citep{Flaherty_2015_hd16_weakturb}. However, this has the additional difficulty of determining surface density and temperature structure parameters \citep[e.g.,][]{Hughes_2009_turb, Guilloteau_2012_turb_dmtau}, which are hard to constrain \citep[see][and references within]{Miotello_ppvii_2023}. Models of CO emission must also account for freeze-out and desorption processes that may be relevant \citep{Oberg_2015_CO_desorp}, adding further complexity in this kind of analysis to obtain turbulence \citep{Flaherty_2020_DMTau}.

Analyzing CO isotopolog data in IM Lup using these complex models also results in high nonthermal motions associated with turbulence of $\mathcal{M}$ = 0.18-0.3 (Flaherty et al. in prep). Within our errors (see Figure 7) these values are consistent with our determination of $\mathcal{M}$ = 0.4-0.6; however, they are a factor of 2 lower. With only one system where it is possible compare both methods, we are not able to conclude if this is a systematic difference or how it would affect the determination of nonthermal motions in other disks. We note that previous best-fit disk radiative transfer models used for studying turbulence in HD\,163296 show strong differences with the disk vertical structure that is more recently constrained by \citet{Paneque_2023_vert}. \citet{Flaherty_2017_DCOp} predict that the expected location for CO ($J=2-1$) in HD 163296, based on their models used for setting upper limits on the turbulence, is $z/r < 0.1$ \citep[see Fig. 16 of][]{Flaherty_2017_DCOp}. Direct observational measurements, however, trace the CO ($J=2-1$) emission surface in HD\,163296 at $z/r\sim$0.3 \citep{MAPS_Law_Surf, Paneque_2023_vert}, indicating that some of the underlying model assumptions may be incorrect and highlighting the need for high resolution data as well as the difficulty and possible errors of such methods. 

Another approach of directly determining the thermal structure and turbulence is to obtain the temperature using multiple transitions, which is particularly useful for face-on disks where the vertical dimension is not accessible. This has been applied to the TW Hya disk to constrain an upper limit on turbulence of $\mathcal{M}\sim$0.1 \citep{Teague_2016_turb_TWHya, Teague_2018_turb_nonlte}. This method is complementary to our approach, as it relies on direct observational constraints rather than physical models; however, it carries the uncertainty on the exact emission height of different transitions.

\section{Conclusions}

We have presented new Band 7 observations at high spectral and spatial resolution of CN $N=3-2$ and C$_2$H $N=4-3$ in IM Lup. The emission from both tracers seems to be optically thin and comes from a vertically elevated layer colocated with $^{13}$CO $J=2-1$. This is the first time C$_2$H is seen to be vertically elevated as predicted by thermochemical models.

Through the use of a parametric model, we traced the emission morphology and constrained the linewidth of CN, which we then decomposed into its thermal and nonthermal components. Assuming that the nonthermal component corresponds to turbulent broadening, we conclude the following:
\begin{enumerate}
    \item IM Lup presents strong turbulence in the upper layers ($z/r \sim$0.25), as traced by CN emission. In this layer and beyond radial distances of 150\,au, turbulent velocities are between $\mathcal{M}$ = 0.4 and 0.6 ($\alpha\sim$10$^{-1}$).
    \item Combined with constraints of low turbulence in the midplane \citep{Franceschi_2023_imlup_turbdust, Jiang_turb_dust}, IM Lup shows evidence for a vertical gradient of turbulence, with higher values in the upper layers. This is a key prediction of MRI \citep{Simon_2015}, which has been proposed to be ongoing in IM Lup due to its ionization structure \citep{Seifert_2021_MRI_imlup}. 
    \item The use of optically thin tracers, with emission constrained to a localized vertical region, together with the temperature profiles extracted from CO isotopologs can be leveraged to directly measure turbulence through simple parametric models of the line intensity, without the uncertainties imposed by the density and temperature structure of protoplanetary disks.
\end{enumerate}

Further use of ALMA's exquisite spectral and spatial resolution will allow us to expand this study to other sources and directly confirm if IM Lup is an exception, or if other disks have turbulent motions comparable to the sound speed in the upper layers.

\begin{acknowledgements}
This paper makes use of ALMA data from project \#2019.1.01357.S. 
ALMA is a partnership of ESO (representing its member states), NSF (USA), and NINS (Japan), together with NRC (Canada),  NSC and ASIAA (Taiwan), and KASI (Republic of Korea), in cooperation with the Republic of Chile. The Joint ALMA Observatory is operated by ESO, AUI/NRAO, and NAOJ. 
Astrochemistry in Leiden is supported by the Netherlands Research School for Astronomy (NOVA), and by funding from the European Research Council (ERC) under the European Union’s Horizon 2020 research and innovation programme (grant agreement No. 101019751 MOLDISK).
This project has received funding from the European Union's Horizon 2020 research and innovation programme under the Marie Sklodowska-Curie grant agreement No 823823 (DUSTBUSTERS). 

\end{acknowledgements} 

\bibliographystyle{aa}
\bibliography{imlup_turb.bib}

\begin{appendix}

\section{Radial features}

\begin{figure}[h!]
   \centering
   \includegraphics[width=\hsize]{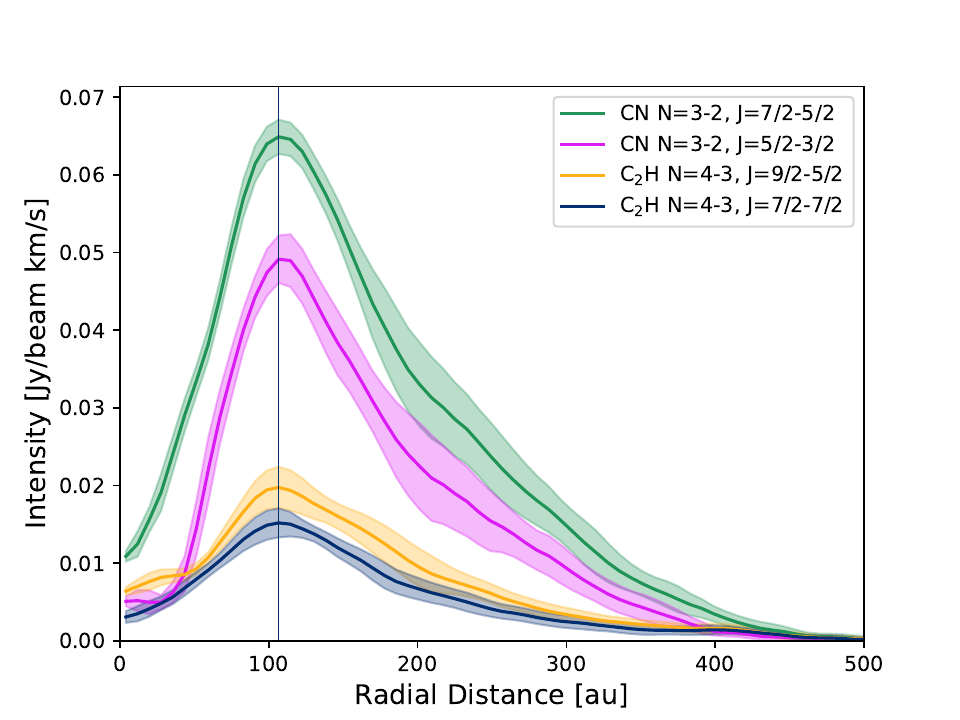}
      \caption{Azimuthally averaged radial intensity profiles for each of the observed hyperfine groups. The solid line traces the average value and the shaded region the dispersion within each radial bin. The vertical black line indicates the peak of emission, which coincides for all transitions at $\sim$106\,au.
              }
         \label{radial_prof}
\end{figure}

The integrated intensity emission maps (moment 0) show a distinct radial feature for all the studied hyperfine groups (Fig. \ref{imlup_emission}). We compute the azimuthally averaged intensity profiles from these maps by accurately deprojecting using the best-fit vertical surfaces and the function radial\_profile from GoFish \citep{gofish}. Only emission from within a 60$^{\circ}$ wedge along the semimajor axis is considered, to avoid deprojection effects from emission close to the minor axis \citep{MAPS_Law_radial}. Our results are shown in Figure \ref{radial_prof}, where all tracers display a ringed structure peaking at $\sim$106\,au.

\begin{figure*}[h!]
   \centering
   \includegraphics[scale=0.7]{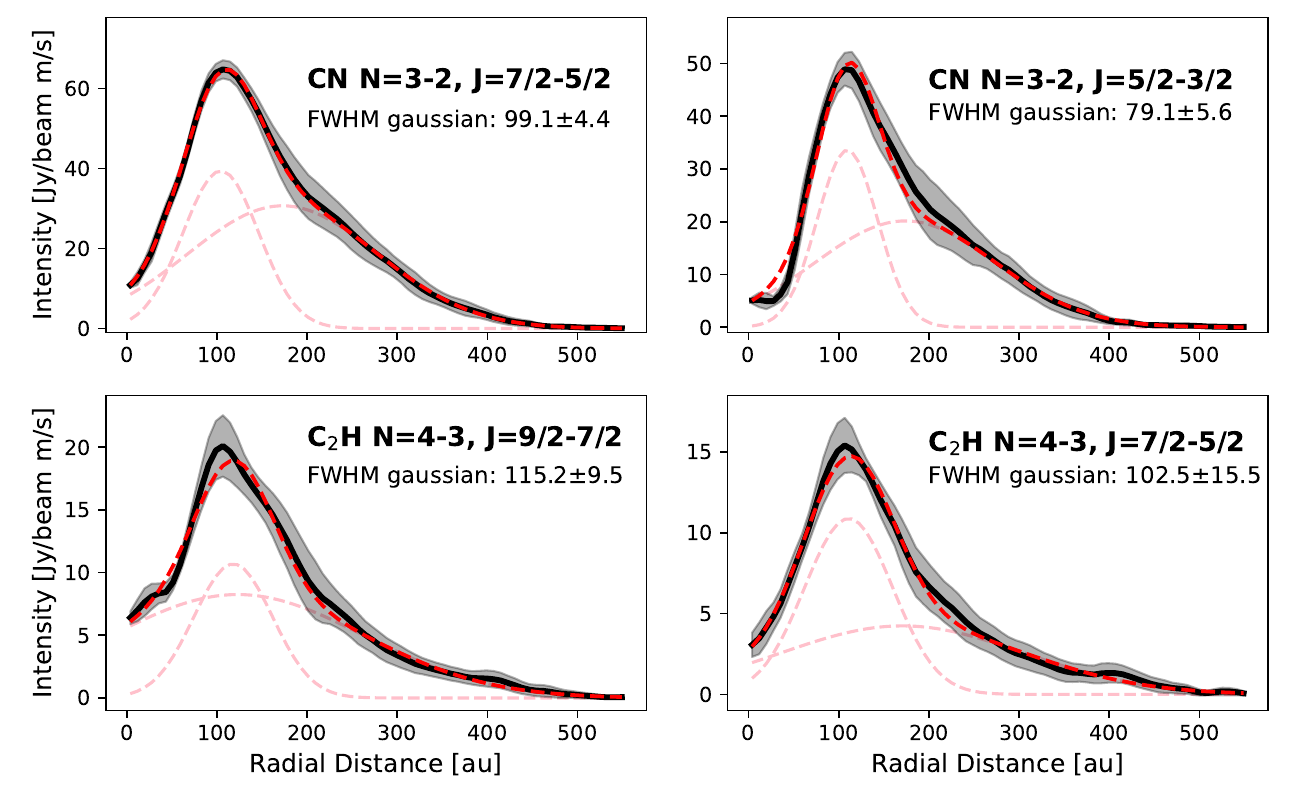}
      \caption{Gaussian fits to the radial intensity profiles of each transition. The data are shown in black, pink dashed lines trace the two individual Gaussian components, and the dashed red line presents the model curve, obtained by summing both Gaussians. The FWHM of the inner Gaussian, which traces the ring component, is indicated for each panel. 
              }
         \label{fits_gauss_ring}
\end{figure*}

To characterize the width and any azimuthal variations of the ring feature, a set of two Gaussians are fitted to the radial profiles as done in \citet{MAPS_Law_radial}. A broad component Gaussian is used to trace the emission at further out radii and a second Gaussian traces the ring structure. Figure \ref{fits_gauss_ring} shows the fitted Gaussians and FWHM value of the ring-component Gaussian. We used a Gaussian function such that

\begin{equation}
    I = I_0 \, \mathrm{exp}\left[ -\frac{1}{2} \frac{(R-R_0)^{2}}{\sigma^2} \right] 
,\end{equation}where $I$ is the measured intensity, $R$ the radial distance, $I_0, R_0$ and $\sigma$ are the fitted parameters and the FWHM = 2$\sqrt{2\,\mathrm{log}(2)}\,\sigma$. While the peak of emission was coherent across different hyperfine groups, the ring width has slight variations. Overall the feature shows an average width of $\sim$89\,au in CN and $\sim$109\,au in C$_2$H. 

\begin{figure*}[h!]
   \centering
   \includegraphics[scale=0.6]{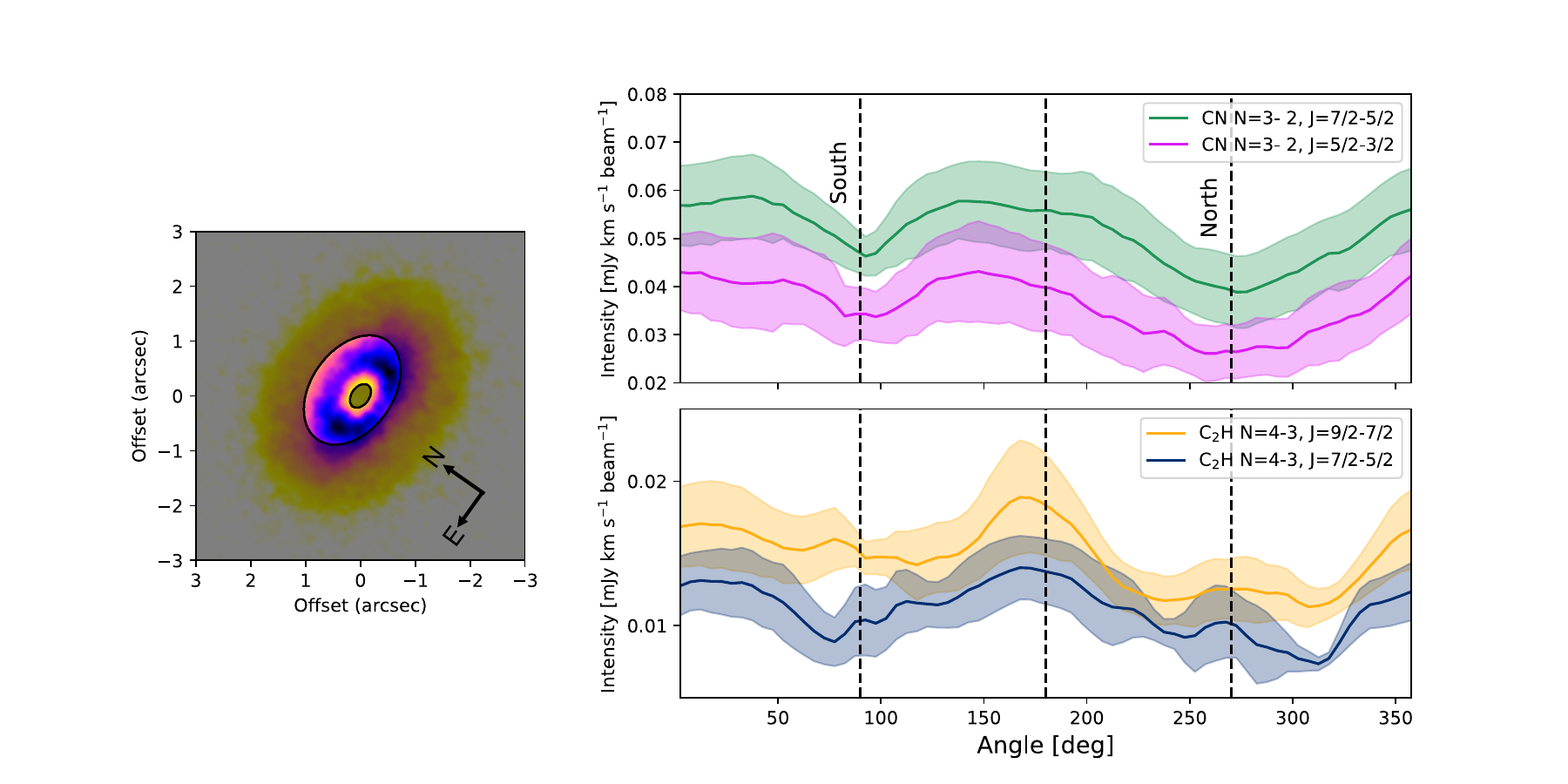}
      \caption{Azimuthal emission curves along the ring for each transition. The solid line indicates the mean values within each azimuthal bin and the shaded region the dispersion. The panel to the left shows the radial limits used to extract the ring emission.
              }
         \label{ring_intensity}
\end{figure*}

Considering the broadest value of $\sim$115\,au we calculate the averaged intensity throughout the angular extent of the ring, using azimuthal bins of 5$^{\circ}$. Figure \ref{ring_intensity} shows the results. CN displays a symmetric ring across azimuth, with slight variations along the semiminor axis likely due to projection effects. C$_2$H has a lower S/N and may show tentative asymmetries; however, the quality of the data is not sufficient to characterize them in detail. 

\section{Channel map analysis}
\begin{figure*}[h!]
   \centering
   \includegraphics[width=\hsize]{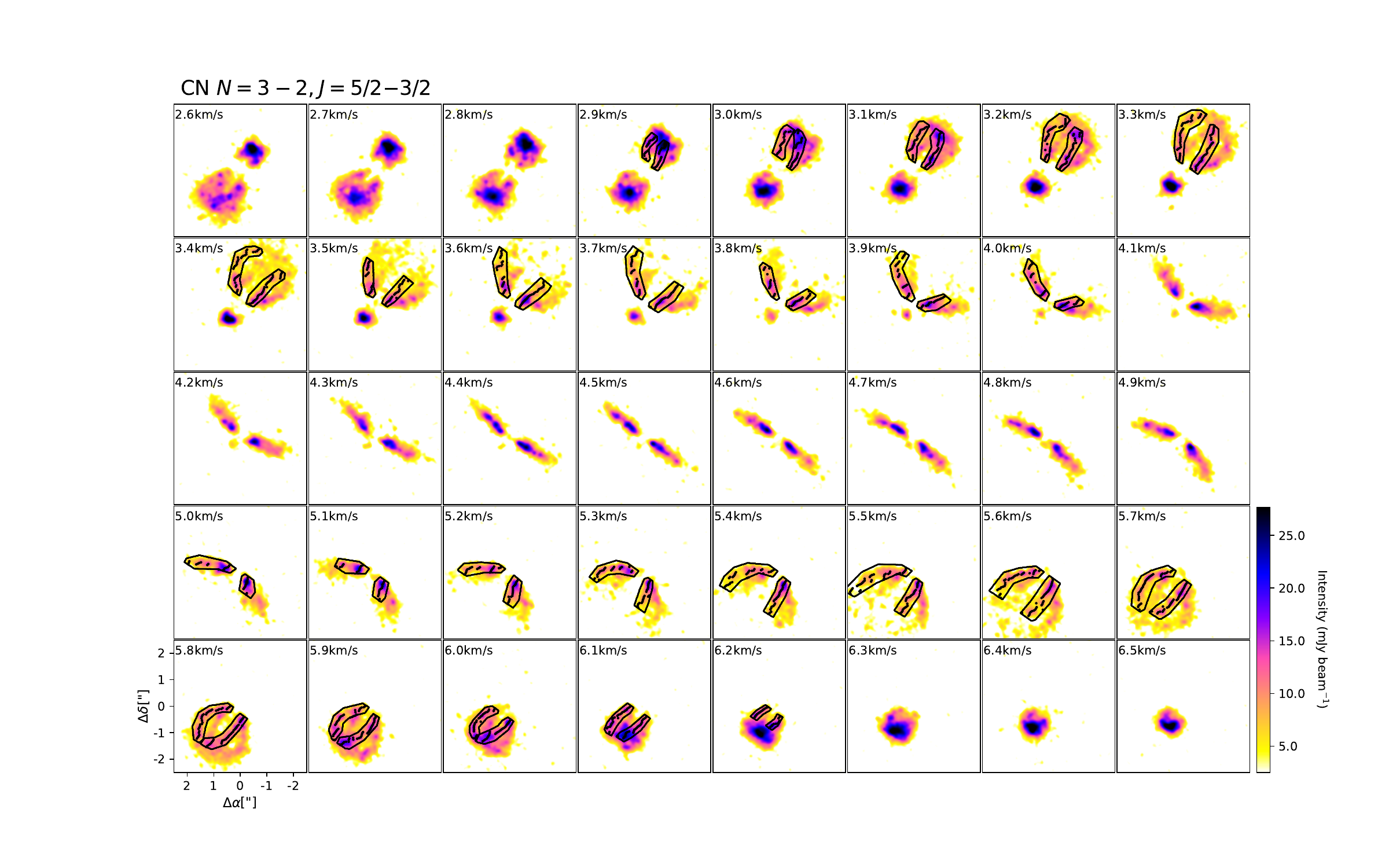}
      \caption{CN $N=3-2$ emission, showing the channel maps used to extract the vertical emission layer with ALFAHOR. The masked regions are indicated in the black contours and black dots within the masks trace the emission maxima.
              }
         \label{channels_cn}
\end{figure*}

\begin{figure*}[h!]
   \centering
   \includegraphics[width=\hsize]{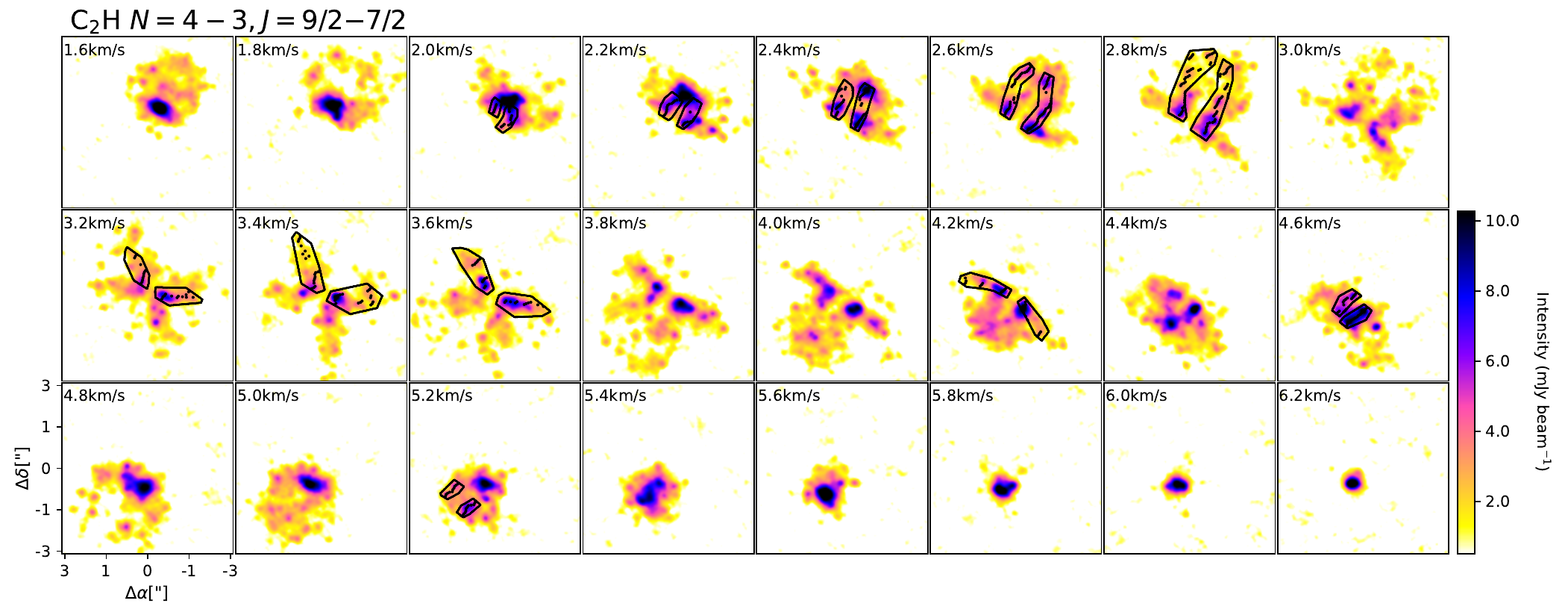}
      \caption{Same as Fig. \ref{channels_cn} but for C$_2$H $N=4-3$ emission.
              }
         \label{channels_c2h}
\end{figure*}

To extract the vertical location of the emission layer, we traced the emission maxima from the channel map emission as explained in section 3.1. The channels, masks, and extracted emission maxima are presented in Figures \ref{channels_cn} and \ref{channels_c2h} for CN and C$_2$H, respectively.

\section{DISCMINER best-fit parameters}

Table \ref{table_params_dm} indicates the best-fit parameters obtained by DISCMINER for the CN emission model.

\begin{table}[h]
\def\arraystretch{1.5}
\setlength{\tabcolsep}{6pt}
\caption{DISCMINER best-fit parameter values.}
\label{table_params_dm}      
\centering
\begin{tabular}{l c c c}       
\hline\hline                
  Attribute & Parameter & Unit & Value \\
\hline\hline

\multirow{4}{*}{Orientation} & $i$& [$^{\circ}$] & 50.53 \\
            & PA & [$^{\circ}$] & 143.76 \\ 
            & $x_c$ & [au] & -1.9 \\
            & $y_c$ & [au] & 1.3\\
\hline
\multirow{2}{*}{Velocity}    & $M_*$  &[M$_{\odot}$] & 1.22\\
            & $v_{\mathrm{sys}}$ & [km\,s$^{-1}$]&  4.49\\
\hline

\multirow{4}{*}{Upper surface} & $z_0$ & [au] & 18.98\\
                & $p$ & [-] & 1.50\\
                & $R_t$ & [au] & 427.50 \\
                & $q$ & [-] &3.34 \\
\hline

\multirow{4}{*}{Lower surface} & $z_0$ &  [au] &8.35 \\
                & $p$ &[-] &  1.47 \\
                & $R_t$ & [au] &  423.38\\
                & $q$ & [-] & 3.91\\
\hline

\multirow{6}{*}{Peak intensity} & $I_0$ & [Jy\,px$^{-1}$]& 0.69 \\
                & $p$ & [-] & -4.30\\
                & $q$ & [-] & 3.60\\
                & $I_g$ & [Jy\,px$^{-1}$]& 2.42 \\
               & $R_g$ & [au] & 32.74\\
               & $\sigma_{g}$ & [au] & 45.97 \\
\hline
\multirow{3}{*}{Linewidth}      & $L_0$ & [km\,s$^{-1}$]& 0.415 \\
                & $p$ & [-] & -0.205 \\
                & $q$ & [-] & -0.238 \\

\hline \hline               
\end{tabular}
\end{table}

\section{Effect of vertical layer width}

\begin{figure*}[h!]
   \centering
   \includegraphics[scale=0.5]{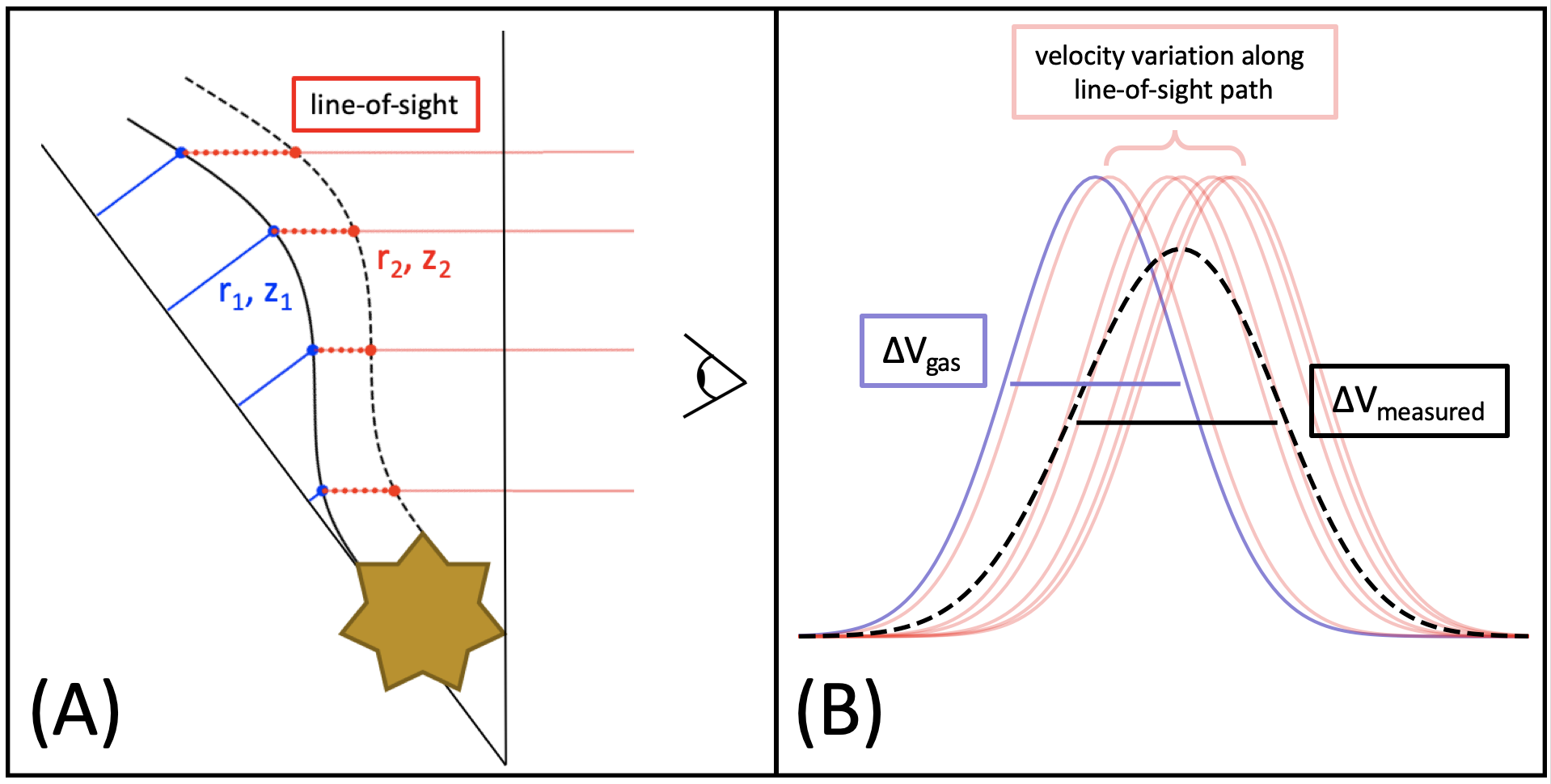}
      \caption{Sketch of the method used to estimate the effect of a nonzero vertical emitting width for CN. Panel (A) shows the geometry of CN emission in IM Lup, considering an inclination of 47.8$^{\circ}$ with respect to the observer. For the vertical structure, the solid curve indicates the best-fit vertical profile of CN emission and the dashed curve the extent considering a fixed vertical width. In each case, the blue dot is the observed parcel of gas at position ($r_1, z_1$) that has to cross the line-of-sight path as indicated by the red line, leaving the zone of CN emission at the red dot location ($r_2, z_2$). For different radial distances, the path varies. Panel (B) shows the velocity variation between the blue dot and the final red dot positions. Each solid-line Gaussian is centered at a determined velocity, depending on their ($r,z$) position along the line-of-sight path. The measured linewidth ($\Delta V_{\mathrm{measured}}$) is taken from the resulting average Gaussian, shown as a dashed black curve and compared to the intrinsic value ($\Delta V_{\mathrm{gas}}$). 
              }
         \label{sketch_width}
\end{figure*}

To estimate the effect of the Keplerian shear due to a nonzero vertical width of the CN emitting layer, we considered the best-fit CN emitting surface as a lower limit and used the geometry of the disk to calculate the line-of-sight path for different vertical widths (see panel A in Figure \ref{sketch_width}). The rotation velocity of a parcel of gas located at a position ($r, z$), in cylindrical coordinates, from the star is given by   

\begin{equation}
    V^{2}_{rot} = \frac{GM_{*}r^{2}}{(r^2 + z^2)^{3/2}},
\end{equation}

\noindent
where $G$ is the gravitational constant and $M_{*}$ the stellar mass. Using this formula, we calculated the velocity for a parcel of material along the line-of-sight path, between the initial ($r_1, z_1$) and final ($r_2, z_2$) positions, as shown in Figure \ref{sketch_width}. The velocity variation is not linear, as it depends on both radial and vertical differences; to be sensitive to these variations, we sampled the line-of-sight path every 1\,au. Using a series of Gaussian profiles with line centers at each of the sampled velocities, we computed the resulting average profile, which is of Gaussian form, and estimated the measured linewidth from it (panel B in Figure \ref{sketch_width}). As the temperature is not expected to change significantly in these upper disk layers for IM Lup \citep{Cleeves_LupusModel_2016}, we assumed a constant intrinsic thermal linewidth in the Gaussian profiles for our estimation. The final linewidth broadening caused by a nonzero emission surface is then calculated by subtracting the intrinsic and measured linewidth values.

\end{appendix}
\end{document}